\documentclass{article}
\usepackage{graphicx}
\usepackage{amsmath}
\usepackage[round]{natbib}
\usepackage{fullpage}
\usepackage{fancyhdr}
\usepackage[ruled,vlined]{algorithm2e}
\usepackage{color}
\usepackage[dvipsnames]{xcolor}
\usepackage{multirow}

\newcommand{\FJP}[0]{\textcolor{Black}{F}}
\newcommand{\NJcP}[0]{\textcolor{Black}{N}}
\newcommand{\RGP}[0]{\textcolor{Black}{R}}
\newcommand{\CLRGP}[0]{\textcolor{Black}{C}}

\newcommand{\FJR}[0]{\textcolor{Red}{F}}
\newcommand{\NJcR}[0]{\textcolor{Red}{N}}

\newcommand{\CLRGR}[0]{\textcolor{Red}{C}}
\newcommand{\SAR}[0]{\textcolor{Red}{S}}

\fancypagestyle{plain}
{
   \fancyhf{}
   \rhead[C]{\textbf{Molecular Biology and Evolution: under consideration for publication}}
}

\begin{document}

\title{Family-joining: A fast distance-based method for constructing generally labeled trees}
\author{Prabhav Kalaghatgi, Nico Pfeifer, and Thomas Lengauer}

\maketitle
\thispagestyle{plain}
\abstract{
The widely used model for evolutionary relationships is a bifurcating tree with all taxa/observations placed at the leaves. This is not appropriate if the taxa have been densely sampled across evolutionary time and may be in a direct ancestral relationship, or if there is not enough information to fully resolve all the branching points in the evolutionary tree. In this paper, we present a fast distance-based agglomeration method called family-joining (FJ) for constructing so-called generally labeled trees in which taxa may be placed at internal vertices and the tree may contain polytomies. FJ constructs such trees on the basis of pairwise distances and a distance threshold. We tested three methods for threshold selection, FJ-AIC, FJ-BIC and FJ-CV, which minimize Akaike information criterion, Bayesian information criterion, and cross-validation error, respectively. When compared with related methods on simulated data, FJ-BIC was among the best at reconstructing the correct tree across a wide range of simulation scenarios. FJ-BIC was applied to HIV sequences sampled from individuals involved in a known transmission chain. The FJ-BIC tree was found to be compatible with almost all transmission events. On average, internal branches in the FJ-BIC tree have higher bootstrap support than branches in the leaf-labeled bifurcating tree constructed using RAxML. $36\%$ and $25\%$ of the internal branches in the FJ-BIC tree and RAxML tree, respectively, have bootstrap support greater than $70\%$. To the best of our knowledge the method presented here is the first attempt at modeling evolutionary relationships using generally labeled trees.
}

\section{Introduction}
Phylogenetic trees are models of evolutionary relationships. The general approach in phylogenetics is to represent evolutionary relationships using bifurcating trees with sampled taxa (represented by so-called labeled vertices) placed at the leaves. Neighbor-joining (NJ) is a popular method for constructing such trees and uses distances between each pair of taxa. Such trees have the maximum number of unsampled ancestors (represented by so-called latent vertices), each ancestor corresponding to a vertex comprising a branching point in the tree. This approach does not allow the labeled vertices to share an ancestor-descendant relationship, and thus may not be appropriate for data sets that have been densely sampled with respect to evolutionary time, for example, genomic sequences of pathogens that have been sampled from individuals who are part of the same transmission chain.

To account for ancestor-descendant relationships \citet{Jombart2011} model evolutionary relationships using a directed acyclic graph in which each edge is directed from a parent to its child. This graph does not contain any latent vertices and is not necessarily connected. In case the graph is disconnected, it is an incomplete representation of the evolutionary relationships among all the labeled vertices.

In related work \citet{Gavryushkina2014} provide a method for constructing so-called sampled ancestor (SA) trees in which labeled vertices come to be placed at internal vertices by contracting terminal branches. The authors do this in a Bayesian inference framework where trees are generated under a model that does not allow labeled vertices to have degree greater than two and, in addition, does not allow latent vertices to have degree greater than three.

Two distance-based algorithms, recursive grouping (RG) and Chow-Liu recursive grouping (CLRG), have been developed by \citet{Choi2010b} for constructing trees which may contain latent vertices with degree greater than two and labeled vertices with degree greater than 0 (so-called generally labeled trees). The authors additionally developed NJc, a method for constructing generally labeled trees by initially constructing a tree using NJ and subsequently contracting all branches that are incident to a latent vertex and are smaller than a preselected threshold. The performance of RG, CLRG, and NJc was compared on simulated data where only the tree topology was varied. In that study, no method clearly outperformed the others.

We developed a distance-based agglomeration method called family-joining (FJ). FJ iteratively identifies, on the basis of a distance threshold, vertices that are in a parent-child or sibling relationship, and introduces latent vertices if required. After inferring all the edges, the branch lengths are estimated using ordinary least-squares (OLS) regression.

RG, CLRG and FJ require the setting of a threshold that determines the model complexity (number of branches) of the output tree. We tested three approaches to threshold selection which minimized Bayesian information criterion (BIC), Akaike information criterion (AIC), and cross-validation (CV) error, respectively.

We compared the performance of FJ-BIC, FJ-AIC, FJ-CV with NJc-BIC, RG-BIC, CLRG-BIC and SA across diverse simulation scenarios. We applied FJ-BIC to an HIV-1 transmission chain data set \citep{Vrancken2014} and checked if the known transmission events were compatible with the FJ-BIC tree. Additionally in the analysis of HIV-1 sequences, we compared the bootstrap support of branches in the FJ-BIC tree and the maximum likelihood tree constructed using RAxML \citep{Stamatakis2006}.

\section{New Approaches}
\subsection{An overview of family-joining}
The family-joining (FJ) method consists of a distance-based agglomeration algorithm for constructing generally labeled trees, and an efficient algorithm for computing ordinary least-squares (OLS) branch lengths. Trees are inferred using the following agglomeration procedure. We initialize a vertex set with all labeled vertices. At each iteration we select from the vertex set, the vertex pair that optimizes the neighbor-joining objective, as defined by \citet{Saitou1987}, see eq. (\ref{eqn:neighborIdentificationStep}) in Materials and Methods. We classify the selected vertex pair as being either parent-child or siblings on the basis of a threshold $\epsilon$, see eq. (\ref{eqn:relationshipTest}) in Materials and Methods. If they are found to be siblings we check if there is another vertex that is the parent of both the siblings. If no such vertex is found, a latent vertex is introduced as the parent of both the siblings. The distance matrix is augmented by adding distances from the newly introduced latent vertex to each of the other vertices, obtained using the formula described in \citet{Studier1988}, see eq. (\ref{eqn:distancesFromLatentVertex}) in Materials and Methods. Rows and columns of the distance matrix corresponding to the children are removed, and the procedure is iterated until a connected graph is obtained. Subsequently, we estimate branch lengths using ordinary least-squares (OLS) regression. For efficient calculation of OLS branch lengths we extended the algorithm by \citet{Bryant1997}, which was designed for leaf-labeled trees, to generally labeled trees. OLS branch lengths may be negative, which has no biological interpretation. To account for this, after estimating the branch lengths, all branches that are shorter than $\epsilon$ and are incident to a latent vertex are contracted. Overall, the procedure is similar to constructing the neighbor-joining tree followed by contracting short branches.

We demonstrate FJ by applying it to a tree-additive distance matrix. A distance matrix is tree-additive if there exists a tree, in which the distance between each pair of labeled vertices is equal to the corresponding sum of lengths of the branches that lie along the unique path between the two vertices. 
\subsubsection{An example using tree-additive distances}
\begin{figure*}
\centering
\textsl{}\includegraphics[width=0.8\textwidth]{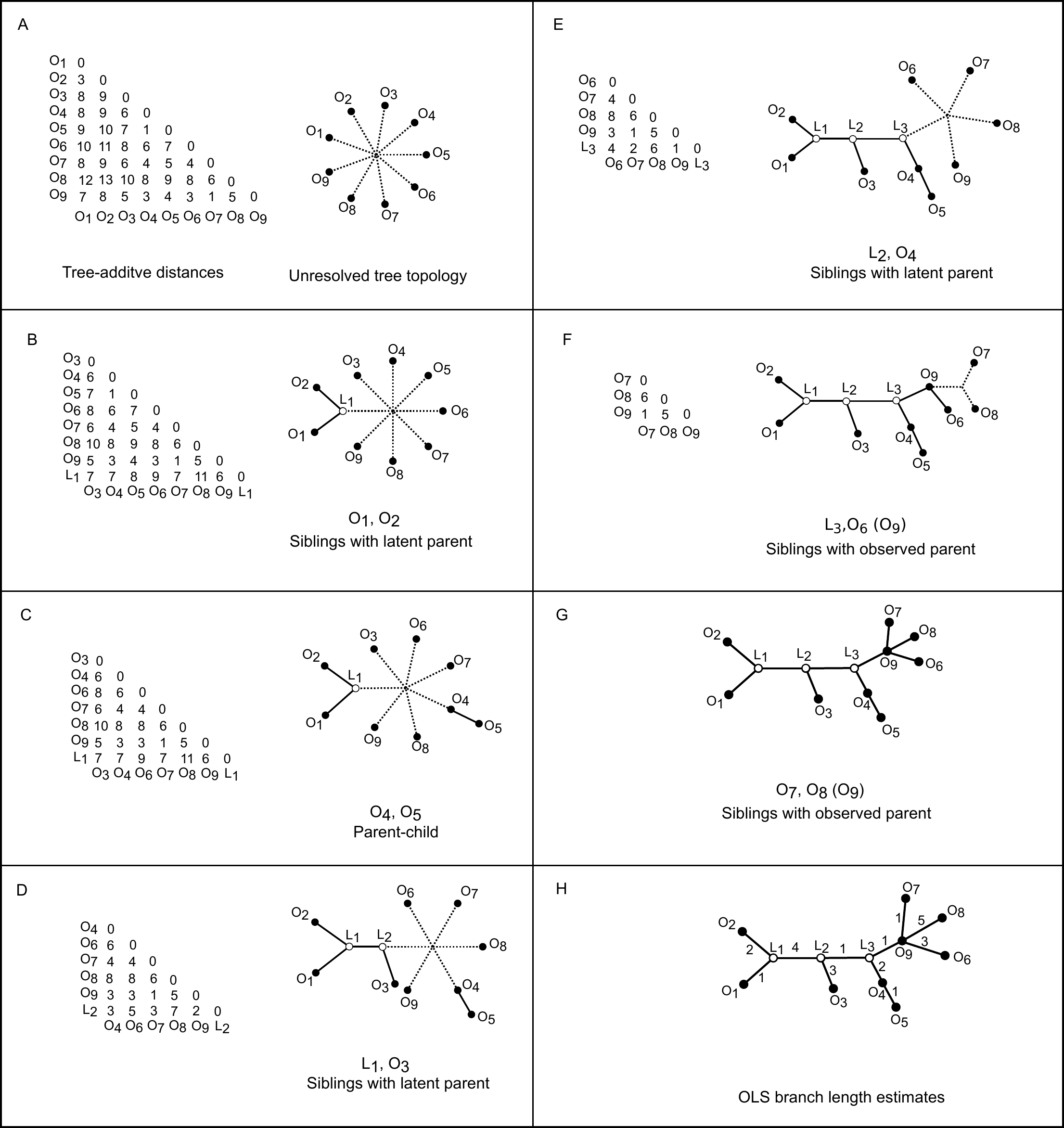}
\vspace*{2 em}
\caption{Panel A: The tree-additive distances used in this example. Labeled vertices are represented by solid circles and latent vertices by white circles with black border. Panels B to G: The agglomeration steps of FJ which identifies the correct tree topology. The edges that are inferred in each agglomeration step are shown as solid lines. The dotted lines connect the labeled and latent vertices that will be used in the next iteration. Panel H: The correct branch lengths estimated using OLS.}
\label{fig:FJ_example}
\end{figure*}

We simulated a generally labeled tree and computed corresponding tree-additive distances. We applied FJ to the resulting tree-additive distance matrix and describe the major steps below. See Fig. \ref{fig:FJ_example} for an illustration. The first iteration identified $O_{1}$ and $O_{2}$ as neighbors that share a sibling relationship. No parent was found for these siblings and a latent vertex $L_{1}$ was introduced. Distances between $L_{1}$ and vertices $O_{3}$ through $O_{9}$ were calculated and the rows and columns corresponding to $O_{1}$ and $O_{2}$ were removed from the distance matrix. Edges were added between $L_{1}$ and $O_{1}$, and between $L_{1}$ and $O_{2}$. The second iteration found $O_{4}$ and $O_{5}$ as neighbors that share a parent-child relationship with $O_{4}$ being the parent. An edge was added between $O_{4}$ and $O_{5}$, and $O_{5}$ was removed from the distance matrix. The following two iterations identified neighbors that are siblings with no parent thus introducing two latent vertices $L_{2}$ and $L_{3}$. The sibling pairs found in the third and fourth iteration are $(L_{1}, O_{3})$ and $(L_{2}, O_{4})$ respectively. The fifth iteration identified $L_{3}$ and $O_{6}$ as siblings, both of which are the children of $O_{9}$. Similarly, the next iteration found $O_{9}$ to be the parent of both $O_{7}$ and $O_{8}$. The final step involved estimating branch lengths using ordinary least-squares. The estimated branch lengths are identical to the corresponding branch lengths in the simulated tree.
\section{Results and Discussion}
\subsection{Simulated data}

Simulated data sets were constructed by varying either the tree type, proportion of labeled internal vertices, type of contracted edge, number of labeled vertices, sequence length or branch length. Each of these parameters is described in detail below. An overview of the parameter settings is provided in Table \ref{tab:simulatedDatasets}.

\begin{table*}[t]
\begin{center}
\caption{Simulated data sets were constructed by varying either the tree type, proportion of labeled internal vertices, type of contracted edge, number of labeled vertices, sequence length or branch length. All settings that were considered for each parameter are shown below. The default setting for each parameter is indicated with $^{*}$.}
\begin{tabular}{|l|c|c|c|c|c|}\hline
\label{tab:simulatedDatasets}
Tree type &&balanced&random$^{*}$&unbalanced&\\\hline
Fraction of latent vertices &0.5&0.37&0.25$^{*}$&0.12&0\\\hline
Contracted edge &\emph{leaf/latent}&\emph{labeled/latent}&\emph{any/latent}$^{*}$&\emph{latent/latent}&\\\hline
Average branch length &0.001&0.004&0.016$^{*}$&0.064&0.256\\\hline
Number of labeled vertices &20&40&80&160$^{*}$&320\\\hline
Sequence length &250&500&1000$^{*}$&2000&4000\\\hline
\end{tabular}
\end{center}
\end{table*}

Three types of binary trees were generated: balanced, unbalanced and random. Unbalanced or ladder-like trees have the largest diameter among all the trees with the same number of vertices. The diameter of a tree is the number of edges that lie on the path in the tree with the maximum number of edges. We chose this tree type because it has been shown that the accuracy of the neighbor identification step (\ref{eqn:neighborIdentificationStep}), which forms a part of FJ, is inversely related to tree diameter \cite{St.John2003}. A balanced tree is complementary to an unbalanced tree and has the smallest diameter possible.

The fraction of latent vertices ranges from zero to $(n-2)/(2n-2)$ where $n$ is the number of labeled vertices. We simulated trees by varying the fraction of latent vertices over this range in four equal steps.

Trees with the desired proportion of labeled vertices were constructed by contracting edges of a binary tree. Depending on the type of simulation experiment, the following edges were contracted: \emph{leaf/latent}, \emph{labeled/latent}, \emph{latent/latent}, and \emph{any/latent}.

For each setting of tree type, fraction of latent vertices, and edge type, we randomly generated corresponding types of binary trees and contracted randomly selected edges of the appropriate type, until the desired fraction of latent vertices was reached. Once the topology was generated, branches were assigned lengths by uniformly sampling numbers between 1 and 100, and scaling them such that the expected branch length was equal to a preselected branch length average. Branch length averages took values of 0.001, 0.004, 0.016, 0.064, and 0.256 subs/site. A vertex was randomly selected as the root and sequences were evolved along the branches according to a GTR+$\Gamma$ model of substitution \cite{Lanave1984}. The parameters of the GTR model were set using estimates from a real data set \cite{Waddell1997}. The parameters shape and scale of the $\Gamma$ model were set to 1 which resulted in a moderate variation of substitution rate across sites. Seq-Gen was used for simulating sequence evolution \cite{Rambaut1997}. Sequence lengths took values of 250, 500, 1000, 2000, and 4000 nt. The number of labeled vertices (taxa) took values of 20, 40, 80, 160, and 320. 

Simulation scenarios were defined by varying each parameter over its range while keeping the remaining parameters fixed at their default setting. The default settings for each parameter are described below. Note that this procedure would result in 22 different parameter combinations. We simulated the corresponding 22 scenarios.

For the categorical parameters \emph{tree type} and \emph{contracted edge type}, the respective default settings were \emph{random} and \emph{any/latent}. These settings were selected as the defaults as they do not restrict the generation of generally labeled trees.

For the continuous parameter, fraction of vertices that are latent, which has a bounded range the midpoint was considered as the default value.

For the following continuous parameters with no upper bound: number of labeled vertices, sequence length, and average branch length, we selected the appropriate range and default settings such that the trend in performance over each parameter range would be apparent.

The default setting for the number of labeled vertices was 160, for the sequence length it was 1000 nt, for the average branch length was 0.016 subs/site.

For each setting of parameter values, 100 trees and corresponding sequences were simulated. For distance-based methods we computed pairwise distances using ML distance estimates under a GTR+$\Gamma$ model, computed using RAxMLv8.2.8 \cite{Stamatakis2014}. For SA which constructs rooted trees we provided sampling times for each labeled vertex. This was done by randomly selected a vertex as the root and defining the sampling time for each labeled vertex as the path length from the root. Note that this method of defining sampling times is equivalent to assuming a strict molecular clock with a clock rate of 1.0. When substitution rates (subs./site/time) follow a strict molecular clock, the distance from the root to each labeled vertex is proportional to the time elapsed since divergence from the root. SA recovers the correct clock rate of 1.0 under the strict molecular clock model in all scenarios except two where the average branch length is very small (0.001 and 0.004; see Supplementary Fig. 3) 

\subsection{Performance metrics}
Precision and recall were used to quantify the accuracy of the various methods at reconstructing the simulated trees. These metrics are defined below.
\begin{align*}
&\text{Precision}(T,\hat{T}) \:\:\:\quad= &\dfrac{|S\cap \hat{S}|}{|\hat{S}|}&\mbox{, and}\\
&\text{Recall}(T,\hat{T}) \!\!\!\!\!\!\qquad\quad= &\dfrac{|S\cap \hat{S}|}{|S|},&
\end{align*} 
where $S$ and $\hat{S}$ are the set of splits corresponding to the simulated tree $T$ and the reconstructed tree $\hat{T}$, respectively. Please note that $S$ contains the split of every branch in $T$, including the terminal branches. Precision and recall range from zero to one. Precision is equal to one only if all the splits in the reconstructed tree are present in the simulated tree. Similarly, recall is equal to one only if all the splits in the simulated tree are present in the reconstructed tree. Please note that we do not report Robinson-Foulds distance, which is popularly used for quantifying reconstruction accuracy, since it would be biased against methods that do not allow polytomies. Each of the reconstruction methods that we tested can achieve the highest and the lowest possible value of recall. Among the reconstruction methods that were compared, only SA can not achieve a precision of one if the simulated tree contains polytomies. We feel that both precision and recall are important measures of reconstruction accuracy.
\subsection{Results of comparative study on simulated data}
We present the results of applying FJ-BIC, NJc-BIC, RG-BIC, CLRG-BIC and SA to all simulated data sets. For methods which have the suffix BIC, we performed threshold selection by minimizing Bayesian information criterion (BIC). For FJ, we also tested FJ-AIC and FJ-CV which optimized Akaike information criterion (AIC), and cross-validation error (CV), respectively. As FJ-AIC and FJ-CV never performed higher than FJ-BIC in any simulation scenario we do not show the results in the main paper. These results are shown in Supplementary Fig. 4. A change in precision or recall is considered to be statistically significant if the corresponding Welch's t-test has a p-value that is smaller than $0.01$. A method is said to have significantly high precision or recall if no other method has significantly higher precision or recall, respectively.

\begin{table*}
\small{
\begin{centering}
\caption{Methods with the significantly highest precision and recall are shown below. All methods that are not significantly worse than the best method are also shown. F, N, R, C, and S stand for FJ-BIC, NJc-BIC, RG-BIC, CLRG-BIC, and SA, respectively. Black and red indicate methods with the highest precision and recall, respectively. The default setting for each simulation parameter is indicated with $^{*}$.}
\begin{tabular}{|c|c|c|c|c|c|}
\hline
\multicolumn{6}{|c|}{Precision, \textcolor{red}{Recall}}\tabularnewline
\hline 
\multirow{2}{*}{Tree type} &  & balanced & random{*} & unbalanced & \tabularnewline
 &  & \FJP, \NJcR, \SAR & \FJP, \FJR, \NJcR, \CLRGR, \SAR & \CLRGP, \CLRGR, \SAR & \tabularnewline
\hline 
\multirow{2}{*}{Contracted edge}  & \emph{leaf/latent} & \emph{labeled/latent} & \emph{any/latent}{*} & \emph{latent/latent} & \tabularnewline
 & \FJP, \NJcP, \FJR, \NJcR, \CLRGR & \FJP, \NJcR & \FJP, \FJR, \NJcR, \CLRGR, \SAR & \RGP, \SAR & \tabularnewline
\hline 
\multirow{2}{*}{Fraction of latent vertices} & 0.5 & 0.37 & 0.25{*} & 0.12 & 0\tabularnewline
 & \NJcP, \SAR & \NJcP, \CLRGP, \SAR & \FJP, \FJR, \NJcR, \CLRGR, \SAR & \FJP, \NJcR, \CLRGR, \SAR & \CLRGP, \CLRGR\tabularnewline
\hline 
\multirow{2}{*}{Average branch length}  & 0.001 & 0.004 & 0.016{*} & 0.064 & 0.256\tabularnewline
 & \CLRGP, \SAR & \FJP, \SAR & \FJP, \FJR, \NJcR, \CLRGR, \SAR & \FJP, \CLRGP, \NJcR, \SAR & \CLRGP, \NJcR, \SAR\tabularnewline
\hline 
\multirow{2}{*}{Number of labeled vertices}  & 20 & 40 & 80 & 160{*} & 320\tabularnewline
 & \FJP, \NJcR, \CLRGR & \FJP, \NJcR, \CLRGR & \FJP, \NJcR, \CLRGR, \SAR & \FJP, \FJR, \NJcR, \CLRGR, \SAR & \FJP, \NJcR, \CLRGR, \SAR\tabularnewline
\hline 
\multirow{2}{*}{Sequence length}  & 250 & 500 & 1000{*} & 2000 & 4000\tabularnewline
 & \FJP, \CLRGP, \SAR & \FJP, \SAR & \FJP, \FJR, \NJcR, \CLRGR, \SAR & \FJP, \NJcP, \CLRGP, \FJR, \NJcR, \CLRGR, \SAR & \FJP, \NJcP, \CLRGP, \FJR, \NJcR,	\SAR\tabularnewline
\hline
\end{tabular}
\label{tab:bestMethods} 
\par\end{centering}
}
\end{table*}\textsl{}

\begin{figure*}[htbp]
\centering
\includegraphics[width=\textwidth]{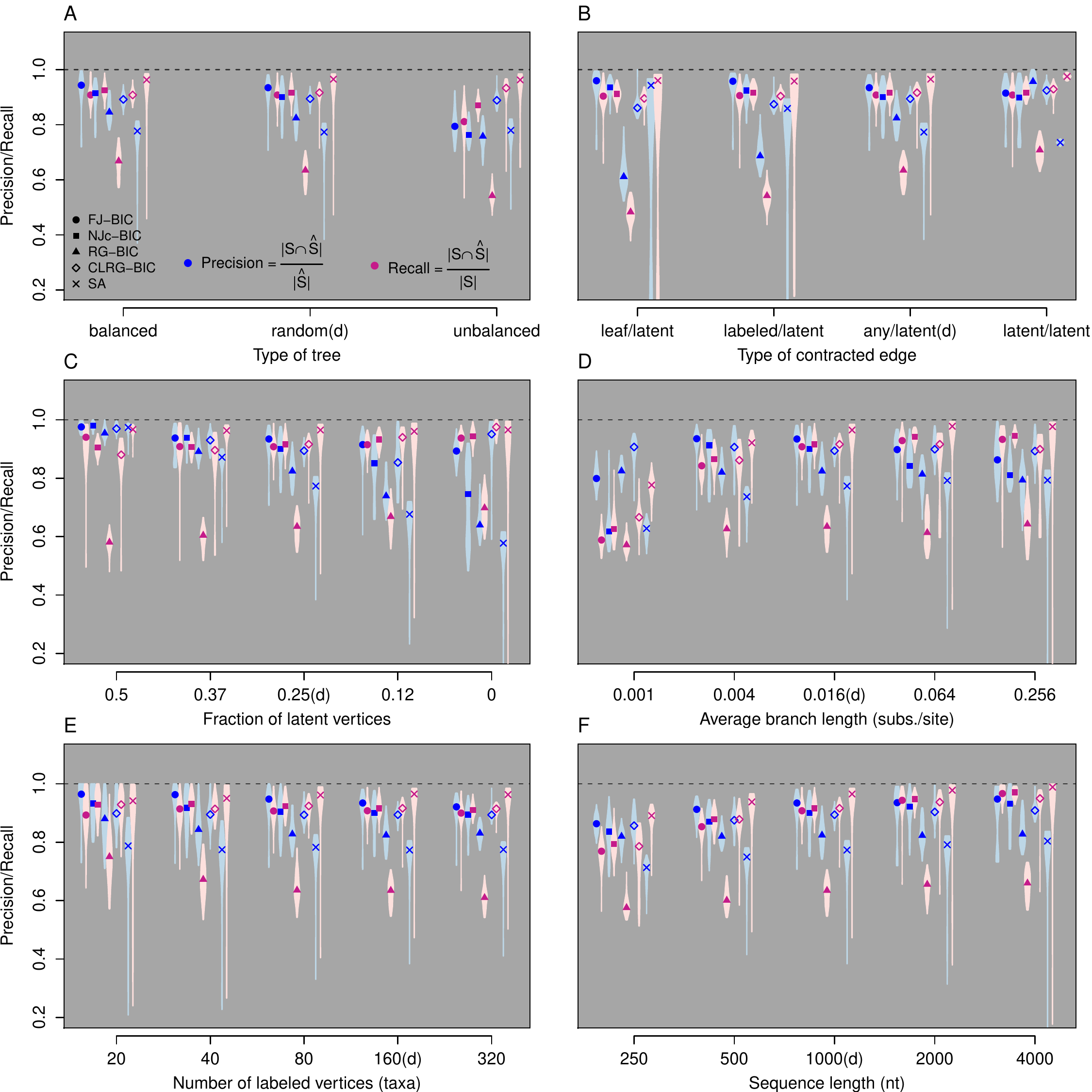}
\vspace*{1 em}
\caption{A comparison of the reconstruction accuracy of all methods in six simulation categories. One parameter (x-axes) was varied in each category. The default parameter settings are denoted as parameterValue(d) on each x-axis. For each parameter setting, 100 data sets were created. Precision is shown in blue and recall is shown in pink.}
\label{fig:precisionRecallAll}
\end{figure*}

\subsubsection*{Tree type}
Both FJ-BIC and NJc-BIC have significantly higher precision and recall on balanced trees than on unbalanced trees. This behavior is expected, since the accuracy of the step of FJ, in which neighbors are identified, is inversely related to tree diameter\cite{St.John2003}. Even on unbalanced trees, which have large diameters, FJ-BIC and NJc-BIC have moderately large (median) precision/recall values of 0.79/0.81 and 0.76/0.87 respectively (see Fig. \ref{fig:precisionRecallAll}A). Similarly RG-BIC performs low on unbalanced trees than on balanced trees, which is in agreement with previous work \cite{Choi2010b}. RG iteratively partitions the entire vertex set into families. Balanced trees and unbalanced trees have $n_\text{leaves}/2$ and two families of size two, respectively. This suggests that RG has a higher error rate for unbalanced trees than for balanced trees. In contrast, CLRG-BIC performs significantly higher on unbalanced trees than on balanced trees with median precision/recall values of 0.89/0.93 and 0.89/0.91, respectively. CLRG constructs the MST and then iteratively applies RG to the neighborhood of each internal vertex. The higher performance of CLRG-BIC on unbalanced trees is most likely due to the MST being topologically close to the unbalanced tree. SA has a median precision and recall of 0.77 and 0.96, respectively, across all tree types. The comparatively lower precision of SA is due to this methods constructing trees in which a labeled vertex can only have up to one descendant and all other internal vertices have degree three. Subsequently this results in trees with excess branches if the true tree contains polytomies.

\subsubsection*{Type of contracted edge}

FJ-BIC has a significantly higher precision than other methods for all types of contracted edges, except \emph{latent/latent}. SA has a high median recall of 0.96 for all types of contracted edges. However the recall values of SA are not significantly higher than those of FJ-BIC if the contracted edge is \emph{leaf/latent}. This is due to a large variance in the performance of SA, quantified with an inter-quantile range of 0.26 (see Fig. \ref{fig:precisionRecallAll}B). SA has high median precision of 0.94 if the contracted edge is \emph{leaf/latent}. Contracting \emph{leaf/latent} edges results in trees in which a labeled vertex can have up to one descendant and all other internal vertices have degree three. The high performance of SA in this category is because these are the same type of trees which SA samples when optimizing tree topology. SA has lower performance when any other edge type is contracted. RG-BIC and CLRG-BIC have significantly higher precision and recall if \emph{latent/latent} edges are contracted, when compared to precision and recall for other edge types.

\subsubsection*{Fraction of vertices that are latent}
For leaf-labeled trees which have a maximal fraction (0.5) of latent vertices, all methods have a median precision higher than 0.95 (see Fig. \ref{fig:precisionRecallAll}C). In this simulation scenario, with a median recall of 0.97, SA has significantly higher recall than other methods, even though FJ-BIC also has a high median recall of 0.94. In general, precision reduces and recall rises with a decrease in the fraction of latent vertices. FJ-BIC has a median precision and recall that is greater than 0.89 across all settings of fraction of latent vertices. CLRG-BIC has a significantly higher precision and recall than other methods when all vertices are labeled. This is because the CLRG algorithm involves the construction of a MST which should be topologically similar to the completely labeled tree.

\subsubsection*{Average branch length (substitution rate)}
All methods perform badly on trees with short average branch lengths of 0.001 subs/site with median recall smaller than 0.8 each (see Fig. \ref{fig:precisionRecallAll}D). This is because a significant portion of the simulated sequences are identical. Thus, in FJ-BIC, NJc-BIC, RG-BIC, and CLRG-BIC there is a preference for choosing parent-child relationship ovSoundser siblings. CLRG-BIC has significantly higher precision than other methods if branch lengths are either very small or very large. FJ-BIC has high precision if the average branch length is between 0.004 and 0.064. In trees with larger branch lengths there is a high chance that sequences undergo multiple substitutions at the same site. This effect has been termed genetic saturation and results in an underestimation of the true evolutionary distance. Additionally, estimates of large distances are associated with large variance \cite{Hoyle2003} which results in the selection of wrong neighbors in the neighbor-joining step. CLRG-BIC has higher performance for trees with large branch lengths because the input to CLRG-BIC is the MST and the construction of the MST is probably robust to noise in distance estimates. The performance of SA is not greatly affected by long branches.

\subsubsection*{Number of labeled vertices (taxa)}
The performance of all the methods is expected to worsen with increasing number of labeled vertices. RG shows significant change in precision/recall with corresponding median values changing from 0.88/0.75 (5 labeled vertices) to 0.83/0.61 (80 labeled vertices) (see Fig. \ref{fig:precisionRecallAll}E). The change in precision and recall shown by SA is not significant. FJ-BIC and CLRG-BIC show a significant drop in precision but not in recall. Even for trees with 320 taxa, FJ-BIC has high median precision and recall of 0.92 and 0.9 respectively. NJc-BIC shows significant change in both precision and recall with median precision/recall changing from 0.93/0.93 to 0.89/0.91.

\subsubsection*{Sequence length}
The performance of all methods improves with increase in sequence length. For all settings of sequence length, FJ-BIC is among the methods with significantly high precision (see Fig. \ref{fig:precisionRecallAll}F). FJ-BIC is among the methods with significantly high recall for sequences of length 1000 nt to 4000 nt. For all settings of sequence length, SA is among the methods with significantly high recall.

\subsubsection*{Summary of performance}
For the simulations performed at the default parameter settings, the methods listed in order of decreasing median precision are FJ-BIC (0.93), NJc-BIC (0.9), CLRG-BIC (0.89), RG-BIC (0.82), and SA (0.77), and the methods listed in order of decreasing median recall are SA (0.96), NJc-BIC (0.92), CLRG-BIC (0.92), FJ-BIC (0.91) and RG-BIC (0.63). In 15 out of the 22 simulated scenarios FJ-BIC is among the methods with significantly high precision. In 17/22 simulated scenarios SA is among the methods with significantly high recall. In 13/22 simulated scenarios NJc-BIC is among the methods with significantly high recall. FJ-BIC has a median recall that is greater than 0.9 in 16/22 simulated scenarios. The remaining scenarios are (i) trees with 20 taxa (recall of 0.89), (ii) trees in which branches are very short (0.001 and 0.004 subs/site; recall of 0.6 and 0.84 respectively), (iii) unbalanced trees (0.81), and (iv) trees constructed using short sequences (250 and 500 nt; recall of 0.77 and 0.85 respectively).

\subsection{Comparison of time-complexities and run times}
\begin{figure*}[htbp]
\centering
\includegraphics[width=0.6\textwidth]{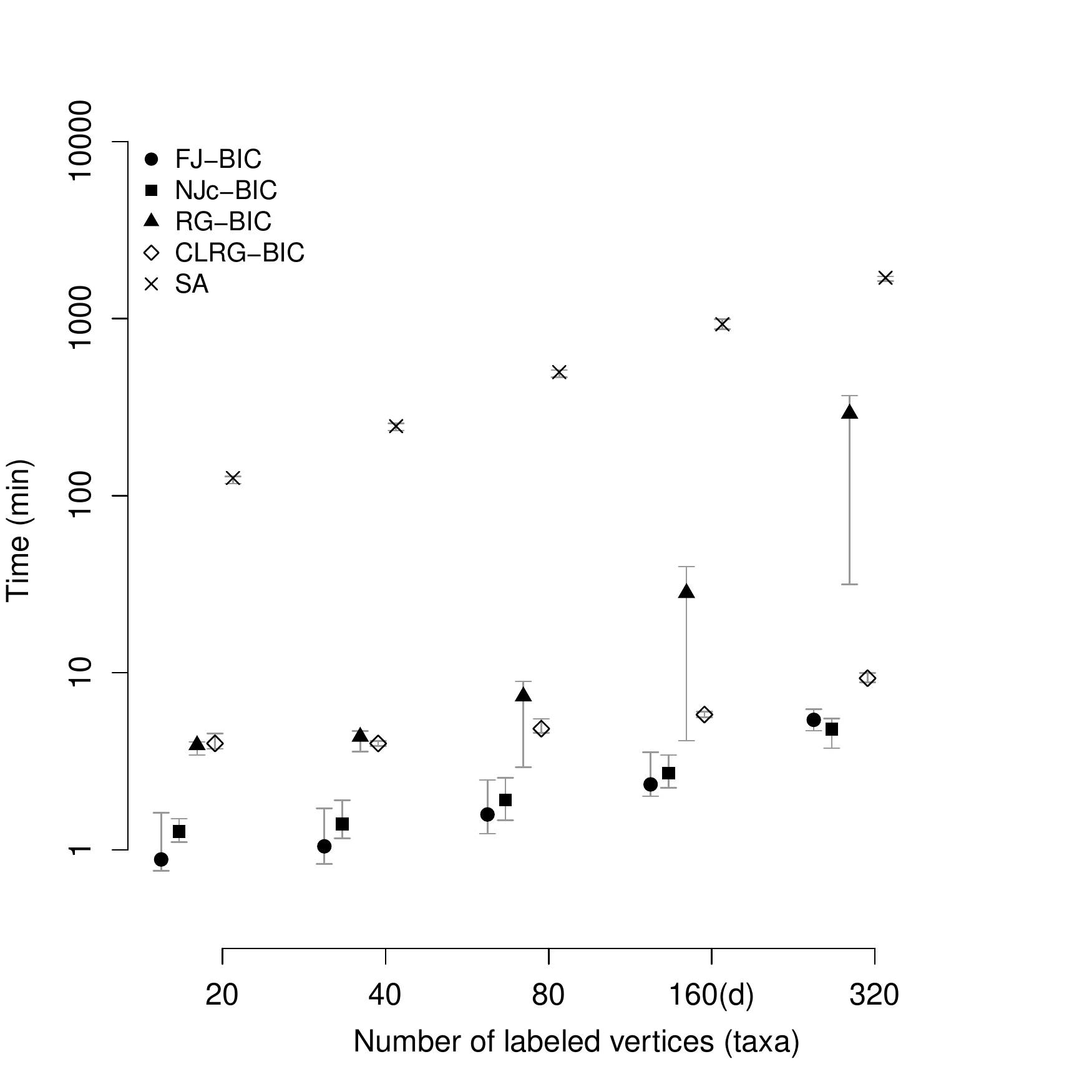}
\vspace*{1 em}
\caption{A comparison of run times of all methods in the scenario where the number of labeled vertices was varied. Run times are shown on a log-scale.}
\label{fig:runTimesAll}
\end{figure*}

Clustering methods are deterministic procedures for which we report worst-case run times. Both FJ and NJ run in time $O(n^{3})$. RG runs in time $O(n^{4})$ which makes it infeasible to run on large datasets. CLRG runs in $O(n^{2}\log n + n_{i}\delta^{3}_{\max}(\mbox{MST}))$ where $n_{i}$ is the number of internal vertices of the MST and $\delta_{\max}(\mbox{MST})$ is the largest vertex degree in the MST. Model selection with BIC or AIC requires the repeated optimization of the likelihood function with respect to parameters of the substitution model. Computing the likelihood with Felsenstein's dynamic programming algorithm \cite{Felsenstein1981} takes $O(nA^{2}L)$ time where $L$ is the sequence length and $A$ is the size of the alphabet. $A$ is four for genetic sequences and 20 for protein sequences. We used RAxML for computing and optimizing likelihoods; RAxML is highly optimized for this task. SA performs Bayesian inference by MCMC sampling, a stochastic procedure whose runtime depends on how easily the MCMC chain moves through the space of trees and model parameters.
The observed run times (see Fig. \ref{fig:runTimesAll}) suggest that FJ-BIC and NJc-BIC are the fastest methods for trees containing up to 320 taxa, with both the methods having a median run time of 5.4 and 4.8 minutes respectively. CLRG-BIC took around 9.3 minutes to reconstruct trees containing 320 taxa and showed the slowest growth in run time. RG showed the largest growth in run time taking 4.8 hours for reconstructing trees with 320 taxa. SA was run with MCMC chain-length set to $10^8$ states. SA took around two hours to construct trees containing 20 taxa and 30 hours for constructing trees containing 320 taxa.

\subsection{HIV-1 transmission chain data}
We applied FJ-BIC to a dataset of HIV-1 subtype C \textit{env} gene sequences that were sampled from 11 hosts who are part of a partially known transmission chain \cite{Vrancken2014}. We selected this dataset because it contains sequences from viruses that are evolutionarily closely related. We discarded 31 sequences which had gaps and analyzed the remaining 181 sequences of length 1376 nt. The hosts are labeled $A,B,C,D,E,F,G,H,I,K,\text{ and } L$. Sequences from multiple time points are available for $A,B,C,D,E,\text{ and } H$. The sampling times for all sequences are known. All the pairs of hosts who were involved in a transmission event are known and have been inferred by interviewing the hosts. The direction of transmission is known for all transmission events except for the transmission between $A$ and $B$. 

Additionally we compared the bootstrap support of branches in the FJ-BIC tree with the branches in the maximum likelihood tree constructed using RAxMLv8.2. \cite{Stamatakis2006}. We first identified the most appropriate model of substitution using JModelTest2 \cite{Darriba2012}. A BIONJ tree \cite{Gascuel1997} was constructed with Jukes-Cantor distances and AIC was computed for the following models of substitution: JC, F81, K80, HKY, TrNef, TrN, TPM1, TPM1uf, SYM, GTR. Variants of all substitution models which included a parameter for invariant sites (I) and/or a Gamma model ($\Gamma$) for inter-site rate variation were also tested. GTR+$\Gamma$+I was the best model, i.e., the one with the smallest AIC score. We constructed a tree with RAXML using the original sequence alignment and the GTRCATI model of substitution, which we refer to as the RAxML tree. The CAT model approximates the Gamma model and enables fast computation \cite{Stamatakis2006}.

We inferred a generally labeled tree using FJ-BIC. Pairwise distances were computed using RAxML which included the following steps \cite{Stamatakis2005}. First a maximum parsimony tree was constructed using stepwise addition and the parameters of the substitution model GTR+$\Gamma$ were optimized. The optimized substitution model was used to compute maximum likelihood distances for all sequence pairs. For computing the likelihood of FJ trees at various values of the distance threshold we used RAxML as follows. FJ trees were converted to leaf-labeled trees by replacing each interior labeled vertex with a latent vertex and adding an edge of length zero between the newly added latent vertex and the former interior labeled vertex. This conversion was necessary since RAxML can only handle leaf-labeled trees. We then maximized the likelihood of the converted FJ tree by fixing the tree topology and branch lengths and optimizing the parameters of the substitution model GTR+$\Gamma$. The maximized log-likelihood was used for computing BIC.

The FJ-BIC tree was rooted assuming a strict molecular clock model. We define the optimal position of the root as that position which minimizes the RSS of regressing distances from the root to each labeled vertex against sampling times. We placed the root at the midpoint of each branch and computed the RSS for each branch. We then picked the branch that had the smallest RSS and searched along the branch for the position of the root with the smallest RSS. Subsequently this position was chosen as the root of the FJ-BIC tree.

\subsubsection{Compatibility of the FJ-BIC tree with known transmission events}
\begin{figure*}[htbp]
\centering
\includegraphics[width=0.8\textwidth]{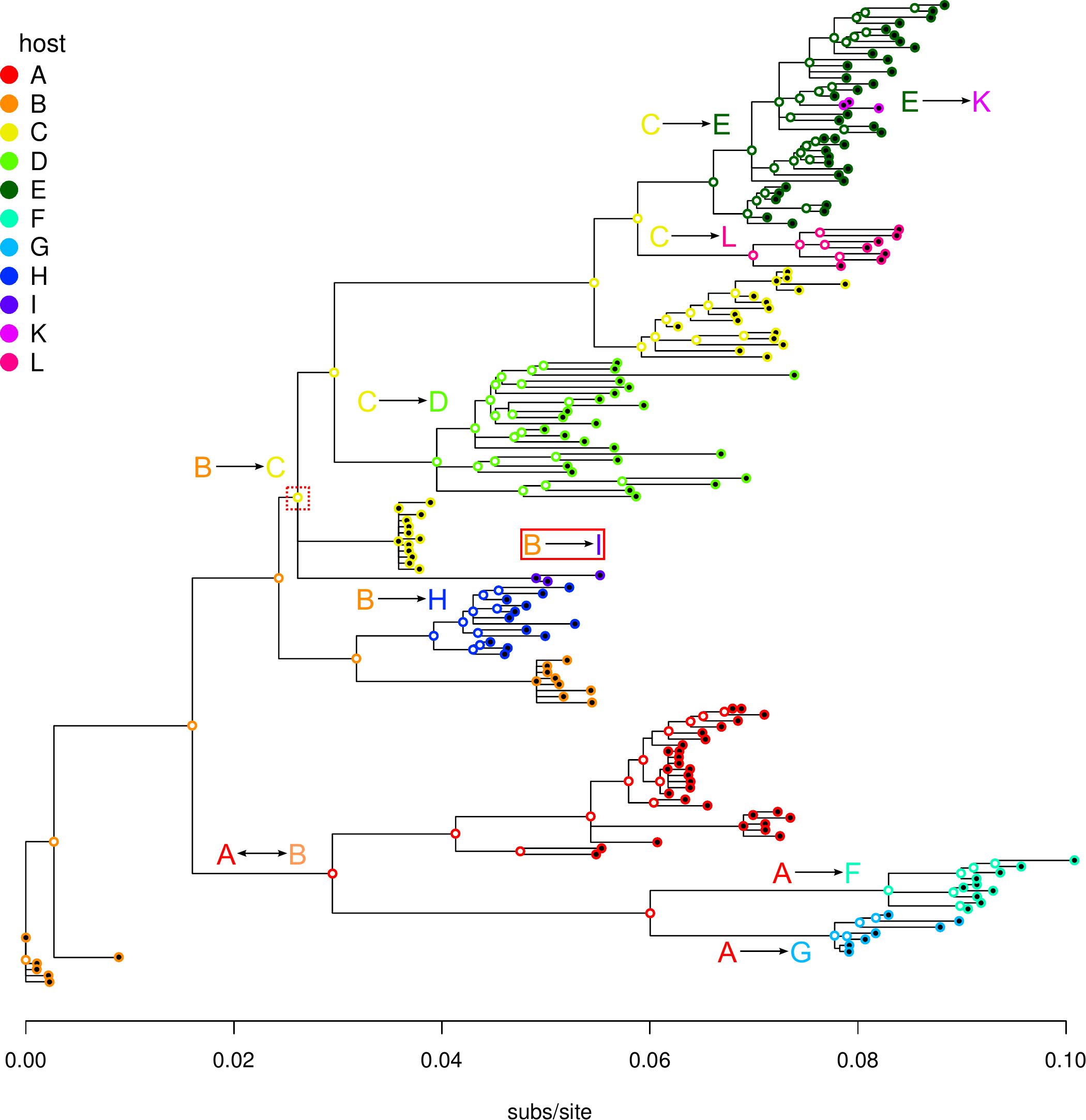}
\vspace{1 em}
\caption{The FJ-BIC tree of 181 HIV-1 \textit{env} gene sequences sampled from hosts involved in a known transmission chain. Each vertex is represented by a circle whose inner color is black if the vertex is labeled and white if the vertex is latent. The outer color of each circle indicates the host of the corresponding vertex. Branches reflecting transmission events have been labeled. 9/10 transmission events are compatible with the FJ-BIC tree. The red box highlights the transmission event $B \rightarrow I$ which is not compatible with the FJ tree.}
\label{fig:FJBICTree}
\end{figure*}

In order to check if the known transmission events are compatible with a rooted tree we needed to label all latent vertices with a host. Latent vertices were visually labeled with hosts using standard maximum parsimony. The labeling that we applied resulted in the minimum possible total cost of 10 (see Fig. \ref{fig:FJBICTree}).
Given a rooted tree with all vertices labeled with a host, we define a transmission event ($X \rightarrow Y$) to be compatible with the tree if there is a directed edge from a vertex labeled $X$ to a vertex labeled $Y$. 9 out of 10 transmission events are compatible with the FJ-BIC tree. The direction of transmission between $A$ and $B$ is not known. The FJ-BIC tree suggests that $A$ was infected by $B$. The branch of the FJ-BIC tree that suggests this transmission event has been labeled with the known transmission event $A \leftrightarrow B$. 8 out the remaining 9 transmission events are compatible with the FJ-BIC tree and branches indicative of these transmission events are labeled in Fig. \ref{fig:FJBICTree}. The transmission event $B \rightarrow I$ is not compatible with the FJ-BIC tree (red solid box in Fig. \ref{fig:FJBICTree}) which may be due to insufficient sampling; Only three sequences were available from host $I$. It is possible that the polytomy present inside the red dotted box in Fig. \ref{fig:FJBICTree} may be resolved if more sequences from $I$ were available, in such a way that the resulting tree would be compatible with the transmission $B \rightarrow I$. 

\subsubsection{Branch support in the FJ-BIC tree and the RAxML tree}

\begin{figure*}[htbp]
\centering
\includegraphics[width=0.42\textwidth]{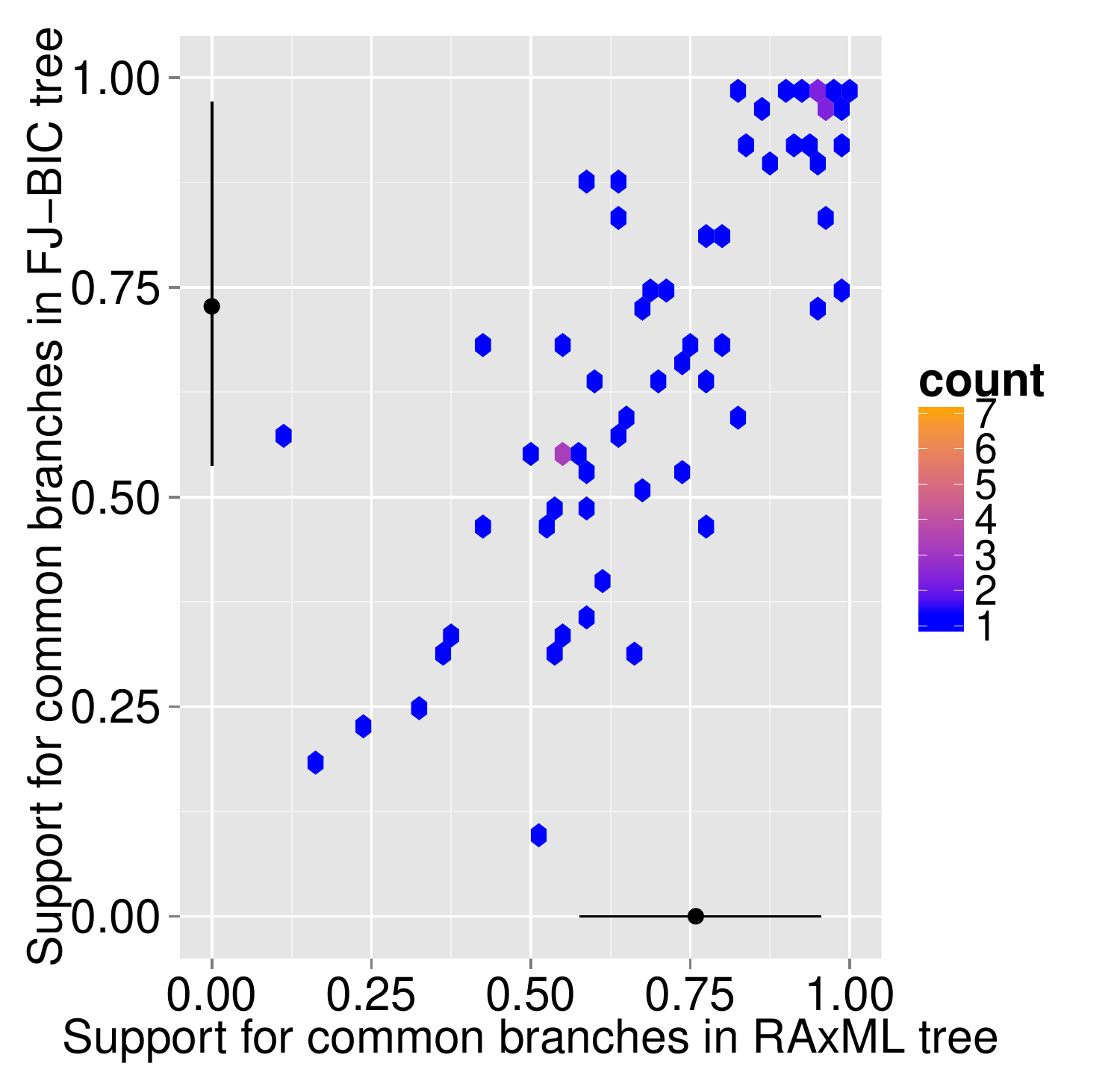}
\includegraphics[width=0.56\textwidth]{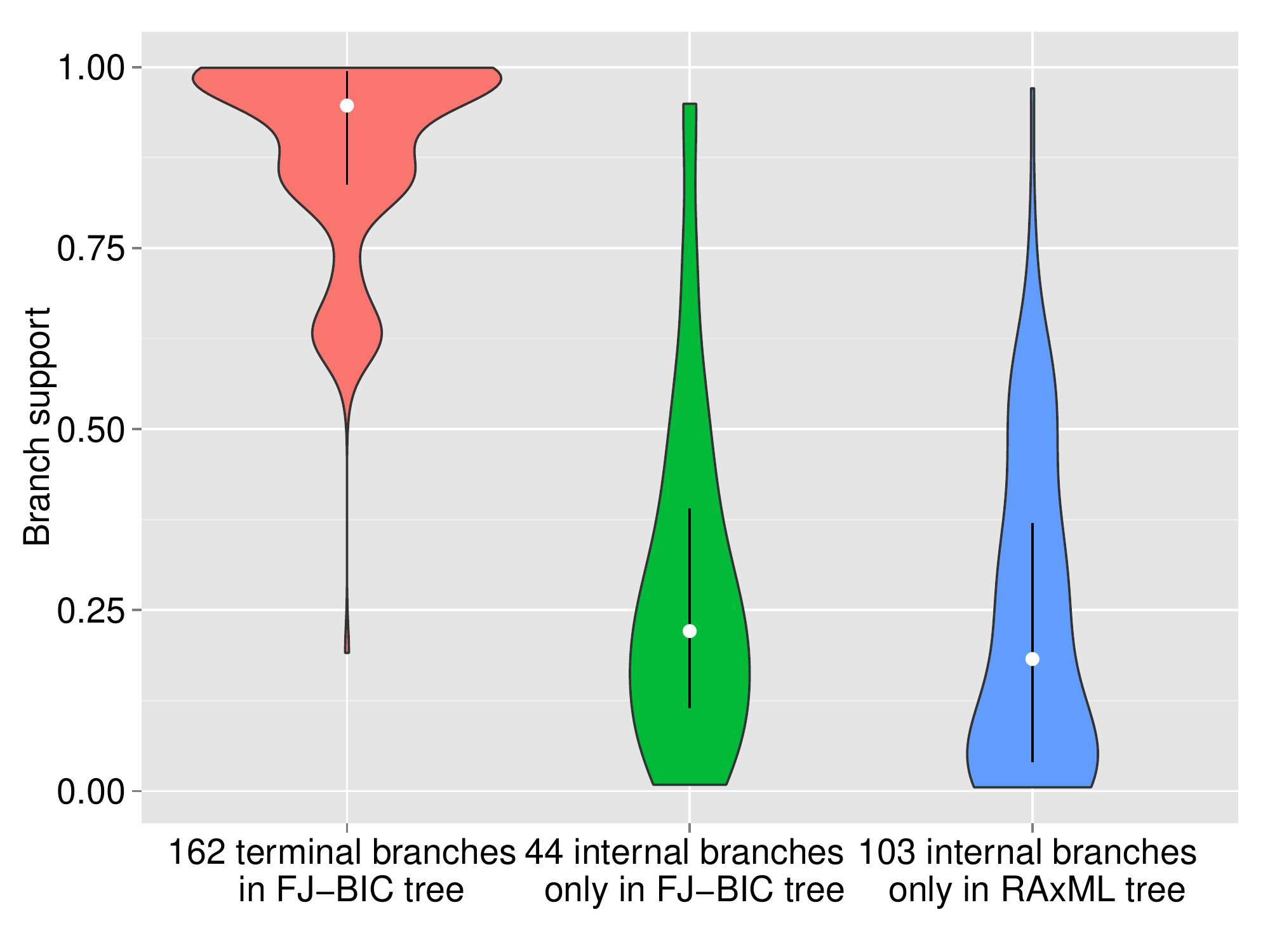}
\vspace*{1 em}
\caption{Left: Comparing the support of common branches in the FJ-BIC tree and the RAxML tree. Right: Supports for branches that are only present in either the FJ-BIC tree or the RAxML tree.}
\label{fig:commonBranches}
\end{figure*}

The bootstrap support of a branch is defined as the number of bootstrap replicate trees that contain this branch. The bootstrap support of branches in the FJ-BIC tree and the RAxML tree were computed using 1000 replicates. Since each labeled vertex is a leaf in all bootstrap RAxML trees, all terminal branches of the RAxML tree trivially have a support of one. The support of a terminal branch in the FJ-BIC tree is not necessarily one.

75 internal branches were common to both the FJ-BIC tree and the RAxML tree. The median(IQR) supports for the common branches were 0.73 (0.43) and 0.76 (0.38) in the FJ-BIC and the RAxML tree respectively. Supports for the common branches in both trees were strongly correlated (Pearson's $\rho = 0.84$, see Fig.\ref{fig:commonBranches}). There are 44 and 103 internal branches that are present only in the FJ-BIC tree and the RAxML tree respectively with lower median (IQR) branch supports of 0.22 (0.28) and 0.18 (0.33) respectively (see Fig.\ref{fig:commonBranches}). The 124 terminal branches in the FJ-BIC tree have a median(IQR) branch support of 0.95 (0.16). 

On average an internal branch in the FJ-BIC tree has a higher support than an internal branch in the RAxML tree. $36\%$ of FJ-BIC branches and $25\%$ of RAxML branches have supports greater than 0.7.

\subsection{Summary and Outlook}
In this paper, we present a fast distance-based agglomeration method called family-joining (FJ) for constructing generally labeled trees. A key feature of the algorithm presented here is its low worst case time complexity, $O(n^{3})$, where $n$ is the number of taxa making it feasible for analyzing large data sets.  For precomputed distances between 320 taxa, FJ-BIC took around 5.4 minutes ($\pm 0.76$) to estimate a tree. At each agglomeration step FJ only adds branches (both internal and terminal) if there is sufficient data to support this move. The algorithm treats short branches as unreliable and identifies an optimal threshold by minimizing test error. We tested two methods, FJ-BIC and FJ-CV error, which minimize BIC and CV error, respectively. When compared with related methods FJ-BIC was best at reconstructing the correct tree across a wide range of simulation settings. FJ-BIC was applied to HIV sequences sampled from individuals involved in a known transmission chain. The FJ-BIC tree was compatible with ten out eleven transmission events. On average, internal branches in the FJ-BIC tree were found to have higher statistical support than internal branches in the tree constructed using RAxML. A method for reconstructing phylogenetic trees with high  precision circumvents the need for a time-consuming bootstrap analysis. To the best of our knowledge the method presented here is the first attempt at modeling evolutionary relationships using generally labeled trees.

\subsection{Summary and Outlook}
In this paper, we present a fast distance-based agglomeration method called family-joining (FJ) for constructing generally labeled trees. A key feature of the algorithm presented here is its low worst case time complexity, $O(n^{3})$, where $n$ is the number of taxa making it feasible for analyzing large data sets.  For precomputed distances between 320 taxa, FJ-BIC took around 5.4 minutes ($\pm 0.76$) to estimate a tree. At each agglomeration step FJ only adds branches (both internal and terminal) if there is sufficient data to support this move. The algorithm treats short branches as unreliable and identifies an optimal threshold by minimizing test error. We tested two methods, FJ-BIC and FJ-CV error, which minimize BIC and CV error, respectively. When compared with related methods FJ-BIC was best at reconstructing the correct tree across a wide range of simulation settings. FJ-BIC was applied to HIV sequences sampled from individuals involved in a known transmission chain. The FJ-BIC tree was compatible with ten out eleven transmission events. On average, internal branches in the FJ-BIC tree were found to have higher statistical support than internal branches in the tree constructed using RAxML. A method for reconstructing phylogenetic trees with high  precision circumvents the need for a time-consuming bootstrap analysis. To the best of our knowledge the method presented here is the first attempt at modeling evolutionary relationships using generally labeled trees.

\section{Materials and Methods}
\subsection{Terminology}
A phylogenetic tree is an edge weighted undirected tree consisting of two types of vertices, labeled vertices (representing observed sequences) and latent vertices (representing unobserved sequences). Sequence information is present only at labeled vertices. Where appropriate, we refer to edges as branches and edge weights as branch lengths. A branch length quantifies the amount of expected change between the sequences corresponding to the respective incident vertices. Branch lengths are usually in units of substitutions per site. Labeled vertices and latent vertices have a minimum degree of one, and three respectively. For a tree consisting of $n$ labeled vertices the number of latent vertices lies between zero and $n- 2$. For trees with a maximal number of latent vertices, all labeled vertices are leaves (degree one) and all latent vertices have degree three. Trees are leaf-labeled if all labeled vertices are leaves, else they are generally labeled. 

A distance matrix $\mathbf{d}$ is tree-additive in a tree $T$ if the distance between each pair of labeled vertices equals the corresponding path length (sum of branch lengths along the unique path between the two vertices) in $T$. Each branch partitions the set of all labeled vertices into two disjoint sets which are referred to as the split of the branch. The two sets of labeled vertices that are present in a split are referred to as the sides of the split. A split is compatible with a tree if there is any branch in the tree such that removing the branch bipartitions the set of labeled vertices as defined by the split. $S(T)$ denotes the set of splits that are defined by a branch in $T$.

A pair of vertices are siblings if both of them are leaves and are adjacent to the same vertex. A vertex pair is in a parent-child relationship if they are adjacent and one of them is a leaf. Thus we call siblings what in the context of the neighbor-joining algorithm is called neighbors.

A rooted tree is a tree with directed edges. In such trees there is one latent vertex (the root) which has indegree zero and outdegree greater than zero. All edges in a rooted tree are directed away from the root.

Edges incident to leaves are referred to as terminal edges while edges incident to internal vertices are referred to as internal edges.
\subsection{Family-joining algorithm}
\begin{figure}[htbp]
\centering
\includegraphics[width=0.35\textwidth]{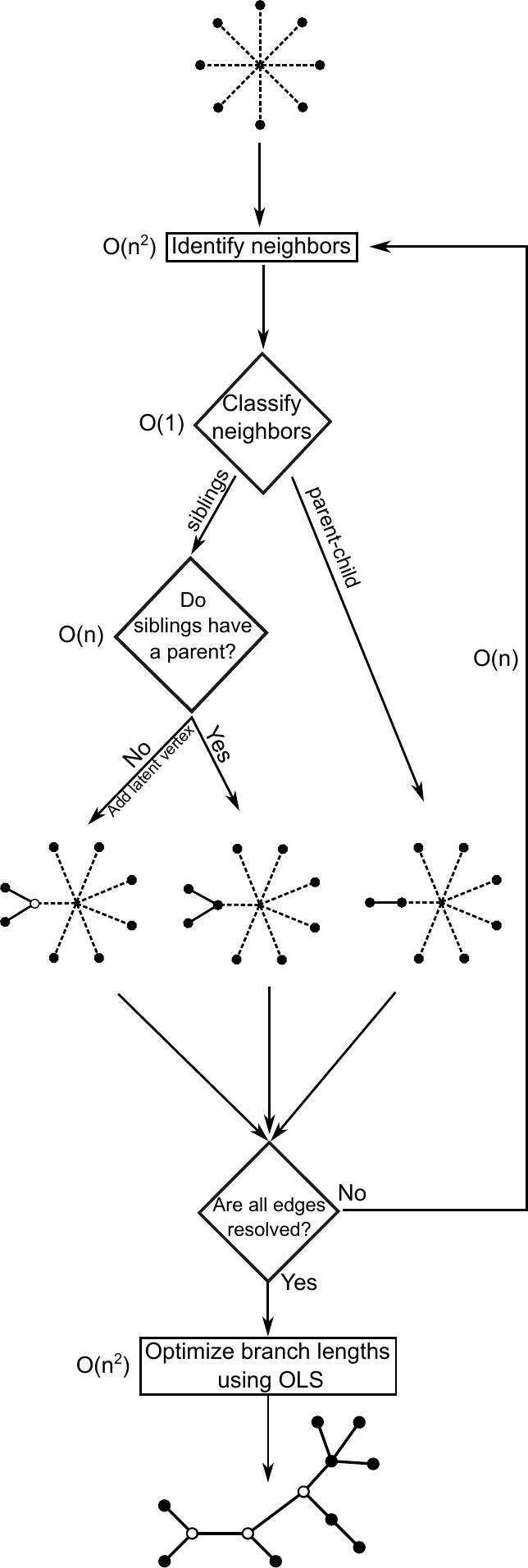}
\vspace*{1 em}
\caption{An illustration of the family-joining algorithm. The main steps have been labeled with their time complexity.}
\label{fig:FJ_illustration}
\end{figure}
Our objective is, given a tree-additive distance matrix $\mathbf{d}$, to infer the respective tree $T_{o}$. $T_{o}$ may be generally labeled and may contain latent vertices with degree greater than three. We assume that all branch lengths in $T_{0}$ are strictly greater than zero. We provide a method which correctly infers $T_{o}$ using entries in $\mathbf{d}$.

Let $\mathcal{T}_{\max}$ be the set of all trees that satisfy the following criteria: (i) their set of leaves includes all the labeled vertices in $T_{o}$, (ii) they have the maximum number of latent vertices, (iii) and $\mathbf{d}$ is the tree-additive distance matrix in every tree in $\mathcal{T}_{\max}$. All splits in $S(T_{o})$ are compatible with every tree in $\mathcal{T}_{\max}$. If this were not true for some tree $T_{\max}$ in $\mathcal{T}_{\max}$ then there would be two branches, $b_o$ in $T_o$ and $b_{\max}$ in $T_{\max}$, such that labeled vertices $\{i, j\}$ and $\{k, l\}$ lie on different sides of $b_0$ and $\{i, k\}$ and $\{j, l\}$ lie on different sides of $b_{\max}$. Applying Buneman's 4-point condition \cite{Buneman1971} would result in the following contradictory inequalities:
\begin{align*}
&d_{ij} + d_{kl} < d_{ik} + d_{jl} \text{ for } b_{0}\\
&d_{ij} + d_{kl} \geq d_{ik} + d_{jl} \text{ for } b_{\max}
\end{align*}

The inequality is strict for $b_{0}$ as all branch lengths in $T_{0}$ are greater than zero.

Thus any tree in $\mathcal{T}_{\max}$ can be constructed as follows. If there is a labeled vertex in $T_0$ with degree greater than one replace this vertex with a latent vertex and add a branch of length zero between the labeled vertex and the newly added latent vertex. If there is a latent vertex with degree greater than three ($v_\text{poly}$) disconnect two randomly selected vertices adjacent to $v_\text{poly}$ and connect them to a new latent vertex with a branch of length zero. Lengths of branches between the newly added latent vertex and each adjacent vertex are the same as the length of the original branch between $v_\text{poly}$ and that vertex. Both of these augmentation operations are performed until all latent vertices have degree 3 and there are no labeled internal vertices.

Applying the neighbor-joining algorithm using distances in $\mathbf{d}$ yields a tree $T_{\it{NJ}}$ with the maximum number of latent vertices such that $\mathbf{d}$ is tree-additive in $T_{\it{NJ}}$. Thus $T_{\it{NJ}}$ belongs to $\mathcal{T_{\text{max}}}$ and consequently neighbors in $T_{\it{NJ}}$ are either parent-child or siblings in $T_{o}$.

NJ is an agglomerative clustering method that identifies, at each step, the pair of vertices to cluster by minimizing the following objective value \cite{Saitou1987,Studier1988}.
\begin{equation}
\label{eqn:neighborIdentificationStep}
(n-2)d_{ij} - \sum_{k\neq i}d_{ik}-\sum_{k\neq j}d_{jk}
\end{equation}
where $n$ is the number of vertices that are yet to be clustered. 

Neighbors $i$ and $j$ can be classified as parent-child or siblings based on the following quantity.
$$\Delta_{ij}=\displaystyle\sum_{k\neq i,j}\dfrac{d_{ji}+d_{ik}-d_{jk}}{2(n-2)}$$
It can be easily shown that:
\begin{equation*}
\begin{aligned}
&\text{ if $i$ is the parent of $j$ }&&\text{then } \Delta_{ij} = 0, \\
&\text{ if $j$ is the parent of $i$ }&&\text{then } \Delta_{ij} = d_{ij}, \\
&\text{ if $i$ and $j$ are siblings }&&\text{then } 0 < \Delta_{ij} < d_{ij} \\
\end{aligned}
\end{equation*}
These criteria are shown to be true in the following statements. If $i$ is the parent of $j$ then the path from $j$ to any vertex $k \neq i,j$, will visit $i$. Thus $d_{jk} = d_{ji} + d_{ik}$, which gives $\Delta_{ij} = 0$ and $\Delta_{ji} = d_{ij}$. If $i$ and $j$ are siblings then $d_{jk} = d_{ju} + d_{uk}$ where $u$ is the vertex adjacent to both $i$ and $j$. Similarly $d_{ik} = d_{iu} + d_{uk}$, which gives $\Delta_{ij} = d_{iu}$. It follows that $0 < \Delta_{ij} < d_{ij}$.

When distances are estimated from sequences we use a threshold $\epsilon$ for classifying the relationship as parent-child or sibling. Specifically $i$ is the parent of $j$ if $|\Delta_{ij}| < \epsilon$. The unordered vertex pair $\{i,j\}$ are said to be in a parent-child relationship if 
\begin{equation}
\label{eqn:relationshipTest}
\min\{|\Delta_{ij}|,|\Delta_{ji}|\} < \epsilon
\end{equation}
The criterion for selecting the appropriate $\epsilon$ is discussed in detail later. When $\mathbf{d}$ is tree-additive any sufficiently small $\epsilon$ can be used for correctly classifying the vertices.

The family-joining algorithm consists of two main parts: GetTreeTopology which infers the tree topology, and GetBranchLengths which estimates the branch lengths. We describe these two steps in detail below.
\subsubsection{Inferring tree topology}
An overview of GetTreeTopology is provided in Algorithm \ref{algo:getTreeTopology}. GetTreeTopology initializes a so-called active vertex set $V_\text{a}$ with the set of all labeled vertices. It then performs agglomerative clustering where the following actions are performed at each step.

The pair $\{i,j\}$ of vertices in $V_\text{a}$ that minimizes (\ref{eqn:neighborIdentificationStep}) is identified. $i$ and $j$ are then classified as parent-child or siblings using (\ref{eqn:relationshipTest}). If $i$ is the parent of $j$, or vice-versa, an edge is added between them and all distances from the child are removed from $\mathbf{d}$. If $i$ and $j$ are found to be siblings then we search for another vertex $k$ in $V_\text{a}$ that minimizes the following quantity. 
\begin{equation}
\label{eqn:deltaParentToChildren}
|d_{ik}+d_{kj}-d_{ij}|
\end{equation}
If $|d_{ik}+d_{kj}-d_{ij}| < 2\epsilon$ then $k$ is the parent of both $i$ and $j$. Corresponding edges are added and all distances from $i$ and $j$ are removed from $\mathbf{d}$. $i$ and $j$ are removed from $V_\text{a}$.
Note that there are alternate criteria for checking if there is a vertex $k$ that is the parent of both $i$ and $j$. One such criterion is to compute
\begin{equation}
\label{eqn:deltaParentToChildren2}
\min\{|\Delta_{ki}|,|\Delta_{kj}|\},
\end{equation}
 and consider $k$ to be the parent of both $i$ and $j$ if $\min\{|\Delta_{ki}|,|\Delta_{kj}|\} < 2\epsilon$. In the simulation study we found that reconstruction accuracy was higher when we used the quantity in eqn. (\ref{eqn:deltaParentToChildren}) as opposed to eqn. (\ref{eqn:deltaParentToChildren2}) (see Supplementary Fig. 4). This is probably because the quantity in eqn. (\ref{eqn:deltaParentToChildren}) is more robust to noise in the estimates of large distances.
If $k$ is not the parent of both $i$ and $j$, a latent vertex $l$ is introduced as the parent of both $i$ and $j$. Corresponding edges are added and distances from $l$ to any vertex $m$ in $V_\text{a}$ other than $i$ and $j$ are calculated using the following estimate by \citet{Studier1988}.
\begin{equation}
\label{eqn:distancesFromLatentVertex}
d_{lm} = (d_{im}+d_{jm}-d_{ij})/2 \qquad\text{for }m\neq i,j
\end{equation}
$i$ and $j$ are removed from $V_\text{a}$ and all distances from $i$ and $j$ are removed from $\mathbf{d}$. Distances from $u$ are added to $\mathbf{d}$ and $u$ is added to $V_\text{a}$.

The agglomeration step described above is repeated until the number of vertices in $V_\text{a}$ is less than four. After each iteration $V_\text{a}$ reduces by either one or two vertices. If $V_\text{a}$ has reached the size three, we check using (\ref{eqn:deltaParentToChildren}) if there are vertices $i$, $j$, and $k$ in $V_\text{a}$ such that $k$ is the parent of both $i$ and $j$. If we find such vertices, corresponding edges are added. Otherwise a latent vertex $u$ is introduced and edges are added between $u$ and the three remaining vertices. If $V_\text{a}$ has reached size two, an edge is added between the two remaining vertices.

GetTreeTopology returns the list of edges of the estimated tree $\hat{T}$. $\hat{T}$ has the same topology as the true tree if distances are additive in the true tree.
\begin{algorithm}
\label{algo:getTreeTopology}
\SetKwInput{Initialize}{Initialize}
\SetKwData{numberOfSingletons}{$s$}
\SetKwFor{While}{while}{do}{}
\KwIn{The distance matrix $\mathbf{d}$, the threshold $\epsilon$, and the labeled vertices $V_{\text{obs}}$}
\Initialize{edge-set $E \leftarrow \emptyset$, $V_{\text{a}} \leftarrow V_{\text{obs}}$}
\While{$|V_{\text{a}}| > 3$}{
From $V_{\text{a}}$ pick $\{i,j\}$ minimizing (\ref{eqn:neighborIdentificationStep})\;
Classify $\{i,j\}$ using (\ref{eqn:relationshipTest})\;
\eIf{$\{i,j\}$ are parent-child}{
Add edge $\{i,j\}$ to $b$\;
Remove child from $V_{\text{a}}$\;
Remove distances from child, from $\mathbf{d}$\;
}{
Remove $i$ and $j$ from $V_{\text{a}}$\;
From $V_{\text{a}}$ pick $k$ minimizing (\ref{eqn:deltaParentToChildren})\;
\eIf{$k$ is the parent of both $i$ and $j$}
{
Add edges $\{i,k\}$ and $\{j,k\}$ to $E$\;
}
{
Introduce vertex $u$ and add it to $V_{\text{a}}$\;
Add edges $\{i,u\}$ and $\{j,u\}$ to $E$\;
Get distances from $u$ (\ref{eqn:distancesFromLatentVertex}), add to $\mathbf{d}$\; 
}
Remove distances from $i$ and $j$, from $\mathbf{d}$\;
}
}
\eIf{$|V_{\text{a}}|=2$}{
$\{i,j\} \leftarrow V_{\text{a}}$; Add edge $\{i,j\}$ to $E$\;
}{
From $V_{\text{a}}$ pick $i,j,k$ minimizing (3)\;
\eIf{$k$ is the parent of both $i$ and $j$}{
Add edges $\{i,k\}$ and $\{j,k\}$ to $E$\;
}{
Introduce vertex $u$\;
Add edges $\{i,u\}$, $\{j,u\}$, and $\{k,u\}$  to $E$\;
}
}
\KwOut{edge-set $E$}
\caption{GetTreeTopology}
\end{algorithm}
\subsubsection{Upper bound on the time complexity of GetTreeTopology}
At first glance it appears that the neighbor identification step requires $\Omega(n^{3})$ time. This can be reduced to $O(n^{2})$ with the observation that the neighbor-joining objective can be reformulated as follows \cite{Studier1988}:
\begin{align}
&(n-2)d_{ij} - R_{i}-R_{j}\nonumber\\
&\mbox{ where } R_{i} = \sum_{k\neq i}d_{ik}\label{eqn:Ri}
\end{align}
From eq. (\ref{eqn:Ri}) it is evident that initializing each row sum $R_{i}$ with the original distances takes $O(n)$ time. Updating each $R_{i}$ after each agglomeration step is done by subtracting distances from children and, if applicable, adding distances to the newly introduced latent vertices. Thus the process of updating each $R_{i}$ takes $O(1)$ time. Additionally, storing all the $R_{i}$ in memory requires $O(n)$ space which incurs very little memory overhead compared to the $O(n^{2})$ space required to store all the pairwise distances. If all distances and row sums are stored in memory then identifying the neighbors takes $O(n^{2})$ time. Note that $\Delta_{ij}$ can also be reformulated for faster computation as follows.
\begin{align*}
\Delta_{ij}&=\sum_{k\neq i,j}\dfrac{d_{ji}+d_{ik}-d_{jk}}{2(n-2)}\\
&=\dfrac{d_{ji}}{2} + \dfrac{(\sum_{k\neq i,j}d_{ik})-(\sum_{k\neq i,j}d_{jk})}{2(n-2)}\\
&=\dfrac{d_{ji}}{2} + \dfrac{(d_{ij}+\sum_{k\neq i,j}d_{ik})-(d_{ji}+\sum_{k\neq i,j}d_{jk})}{2(n-2)}\\
&=\dfrac{d_{ji}}{2} + \dfrac{(\sum_{k\neq i}d_{ik})-(\sum_{k\neq j}d_{jk})}{2(n-2)}\\
&=\dfrac{d_{ji}}{2} + \dfrac{R_{i}-R_{j}}{2(n-2)}.
\end{align*}
Thus, once the neighbors $\{i,j\}$ have been identified, it takes $O(1)$ time to compute both $\Delta_{ij}$ and $\Delta_{ji}$. It takes $O(n)$ time to find the vertex $k$ which minimizes |$d_{ki}+d_{kj}-d_{ij}$|. The overall time-complexity of GetTreeTopology is $O(n^{3})$. The time-complexities associated with the main steps of GetTreeTopology are shown in Fig. \ref{fig:FJ_illustration}.

\subsubsection{Efficient estimation of branch lengths}
Branch lengths $b$ of $\hat{T}$ are estimated by ordinary least squares. This is done by solving $\mathbf{A}b=d$ where $d$ is a column vector containing all those entries of $\mathbf{d}$ that are above or alternatively all those entries of $\mathbf{d}$ that are below the diagonal. $\mathbf{A}$ is the branch incidence matrix of $\hat{T}$ and is constructed as follows. If the $m^{th}$ entry of the $d$ is $d_{ij}$, then
\begin{equation}
\label{eqn:constructBranchIncidenceMatrix}
a_{me} = \begin{cases} 1 & \mbox{if } \mbox{the path from $i$ to $j$ contains $e$} \\ 0 & \mbox{otherwise} \end{cases}
\end{equation}
$\mathbf{A}$ has the dimension $n(n-1)/2 \times |E|$ where $|E|$ is the number of branches in the tree, $n$ is the number of labeled vertices, and $b$ is the vector of branch lengths that we wish to estimate.

The ordinary least squares (OLS) estimate of branch lengths is given by 
\begin{equation}
\label{eqn:OLSEstimate}
\hat{b} = (\mathbf{A}^{t}\mathbf{A})^{-1}\mathbf{A}^{t}d.
\end{equation}

For the estimation of OLS branch lengths we do not make the assumption that distances are tree-additive. For leaf-labeled trees there is a fast $O(n^{2})$ algorithm for computing the OLS branch lengths \cite{Bryant1997}. Any algorithm that estimates OLS branch lengths by performing the matrix operations that are defined in eqn. (\ref{eqn:OLSEstimate}) needs to use all entries of the distance vector, and thus must run in $\Omega(n^{2})$ time \citep{Bryant1998}. Thus the algorithm by \citet{Bryant1997} is time-optimal. We show that this algorithm extends to generally labeled trees. The main steps involved in this computation are computing first $\mathbf{A}^{t}d$ and then $(\mathbf{A}^{t}\mathbf{A})^{-1}\mathbf{A}^{t}d$, each in $O(n^{2})$ time. We describe both of these steps below.

\noindent
Computing $\mathbf{A}^{t}d$

\noindent
The $i^{\mbox{th}}$ entry of $\mathbf{A}^{t}d$, $\delta^{T}_{i}d$, is the sum of all distances between labeled vertices $a$ and $b$ that lie on either side of edge $e_{i}$. $\delta_{i}$ is the $i^{\mbox{th}}$ column of $\mathbf{A}$.
For efficient computation of $\mathbf{A}^{t}d$, edges are visited in order of increasing distance from leaves, keeping track of which edges have already been visited.

We first compute $\delta^{T}_{i}d$ for every terminal edge $e_{i}$ which is defined as follows.
\begin{equation}
\label{eqn:deltaId_terminal}
\delta^{T}_{i}d = \sum_{j,j \neq i} d_{ij}
\end{equation}
Next we compute $\delta^{T}_{i}d$ for every internal edge $e_{i}$ which are visited in the order of increasing distance from leaves. Consider the internal vertex $\alpha$ with only one incident edge $e_{i}$ such that $\delta^{T}_{i}d$ has not been calculated. Let the edges incident to $e_{i}$ be $e_{j_{1}},\ldots,e_{j_{m}}$

Let $C_{i}$ be the side of the split of the edge $e_{i}$ that does not contain $\alpha$. Similarly $C_{j_{k}}$ is the side of the split of $e_{j_{k}}$ that does not contain $\alpha$.

Depending on whether $\alpha$ is labeled or not labeled, $\delta^{T}_{i}d$ is computed as follows:

Case 1: Vertex $\alpha$ is not labeled \cite{Bryant1997}.
\begin{equation}
\label{eqn:deltaId_internal_not_labeled}
\begin{aligned}
\delta^{T}_{i}d &= \sum_{k}\sum_{a \in C_{j_{k}}, b\in C_{i}}\mkern-18mu d_{ab}\\
&=\sum_{k} \delta^{T}_{j_{k}}d -2\sum_{k<l}\sum_{a\in C_{j_{k}},b\in C_{j_{l}}}\mkern-18mu d_{ab}
\end{aligned}
\end{equation}

Case 2: Vertex $\alpha$ is labeled.
\begin{equation}
\label{eqn:deltaId_internal_labeled}
\begin{aligned}
\delta^{T}_{i}d &= \sum_{k}\sum_{a \in C_{j_{k}}, b\in C_{i}} \mkern-18mud_{ab} + \sum_{b \in C_{i}} d_{\alpha b}\\
&=\sum_{k} \delta^{T}_{j_{k}}d -2\sum_{k<l}\sum_{a\in C_{j_{k}},b\in C_{j_{l}}}\mkern-18mud_{ab} - \sum_{k}\sum_{b\in C_{j_{k}}}d_{\alpha b}+ \sum_{b \in C_{i}} d_{\alpha b}
\end{aligned}
\end{equation}

Computing each element of $\mathbf{A}^{t}d$ involves the summation of entries of the distance vector. Since each element of the distance vector is summed over just once, $\mathbf{A}^{t}d$ is computed in $O(n^{2})$ time.

\noindent
Computing $(\mathbf{A}^{t}\mathbf{A})^{-1}(\mathbf{A}^{t}d)$

There is a closed-form solution for the OLS branch length $b_{0}$ of any edge $e_{0}$ which is formulated in terms of the splits, and the elements of $\mathbf{A}^{t}d$, that are defined by $e_0$ and the edges adjacent to $e_{0}$. A description of the branch length formula is given later.

When computing branch lengths, edges can be visited in any order. We derive the branch length formula for an internal edge.

\noindent
\begin{figure*}

  \centering
\includegraphics[width=\textwidth]{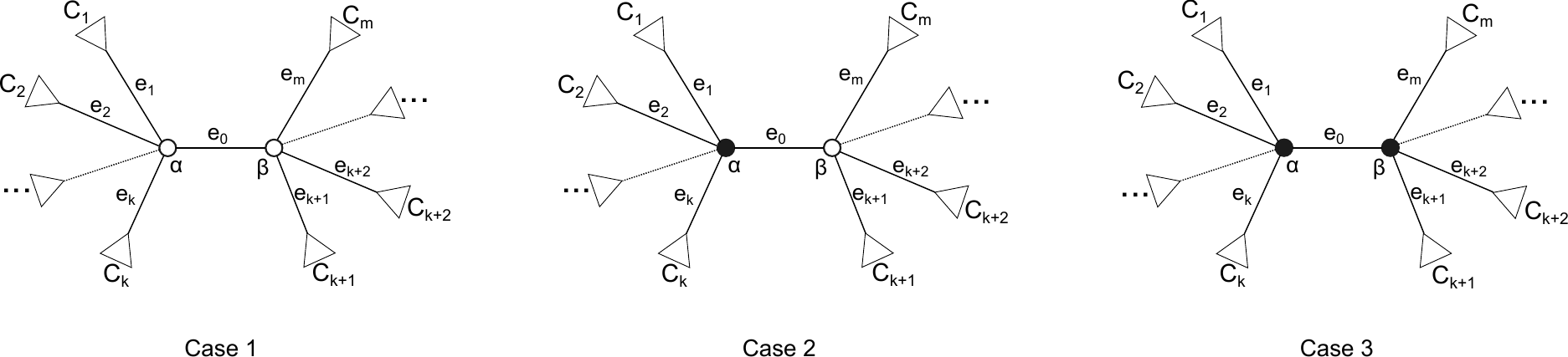}
\vspace*{0.2 em}
\caption{The three cases for the internal edge $e_{0}$. Case 1: Both $\alpha$ and $\beta$ are not labeled. Case 2: Only $\alpha$ is labeled. Case 3: Both $\alpha$ and $\beta$ are labeled. The triangles represent subtrees.}
\label{fig:internalEdges}
\end{figure*}
Consider the internal edge $e_{0}$ shown in Fig. \ref{fig:internalEdges} with adjacent edges $e_{1}, \ldots e_{k}, e_{k+1} \ldots e_{m}$. $e_{0}$ is incident to $\{\alpha,\beta\}$. The respective sizes of the parts of the split defined by $e_{0}$ are $n_{\alpha}$ and $n_{\beta}$

For each edge $e_{i}$ define $P_{i} = \sum_{x\in A_{i}, y \in B_{i}} p_{xy}$ where $A_{i}$ and $B_{i}$ are the parts of the split defined by edge $e_{i}$. Here $p_{xy}$ denotes the length of the path from $x$ to $y$ when branch lengths are determined by OLS. It turns out that $P_{i} = \delta_{i}^{T}d$.

For each edge $e_{i}$ let $C_{i}$ be the side of the split that does not contain $\alpha$ and $\beta$. $n_{i}$ is the cardinality of $C_{i}$. Define
\begin{equation*}
  Q_{i}=\begin{cases}
    \sum_{x \in C_{i}} p_{\alpha x}, & \mbox{if } 1 \leq i \leq k\\
    \sum_{x \in C_{i}} p_{\beta x}, & \mbox{if } k+1 \leq i \leq m\\
  \end{cases}
\end{equation*}

For the case where both $\alpha$ and $\beta$ are not labeled it can be shown that \cite{Bryant1997}
$$\underline{P} = (nI - 2N)\underline{Q} +NU\underline{Q} + b_{0}N\underline{v}$$
where $N$ is the $m \times m$ diagonal matrix with $(n_{1},n_{2},\ldots, n_{m})$ on the diagonal, $I$ is the identity matrix, $\underline{Q} = (Q_{1},Q_{2},\ldots,Q_{m})^{T}$, $U$ is the $m \times m$ matrix of ones, $\underline{v}$ is the vector with $n_{\beta}$ in positions $1$ to $k$ followed by $n_{\alpha}$ in positions $k+1$ to $m$, and $\underline{P} = (P_{1},P_{2},\ldots,P_{m})^{T}$. 

Similarly for the internal edge $e_{0}$
\begin{equation*}
P_{0}=\underline{v}^{T}\underline{Q}+ n_{\alpha}n_{\beta}b_{0}
\end{equation*}
Letting $X = (nN^{-1}-2I+U)$ and substituting $\underline{Q}$ gives the following branch length estimate.
\begin{equation}
\label{eqn:branchLengthEstimate}
b_{0} = \dfrac{P_{0}-\underline{v}^{T}X^{-1}N^{-1}\underline{P}}{n_{\alpha}n_{\beta}-\underline{v}^{T}X^{-1}\underline{v}}
\end{equation}
For cases where only $\alpha$ and both $\alpha$ and $\beta$ are labeled, respectively, the derivation of the above mentioned equations is similar to that described in \citet{Bryant1997} and is provided in the supplementary material.

The formula, eqn. (\ref{eqn:branchLengthEstimate}), for branch length is valid only when $X^{-1}$ exists. \citet{Bryant1997} showed that $X$ is invertible as long as there is at most one zero on the diagonal of the matrix $(nN^{-1}-2I)$. The $i^{th}$ diagonal element is zero if $n_{i}/n = 2$ which occurs if there is an edge where both parts of the split have equal size. Even in generally labeled trees there can be at most one such edge.

\begin{figure}
\centering
\includegraphics[width=0.45\textwidth]{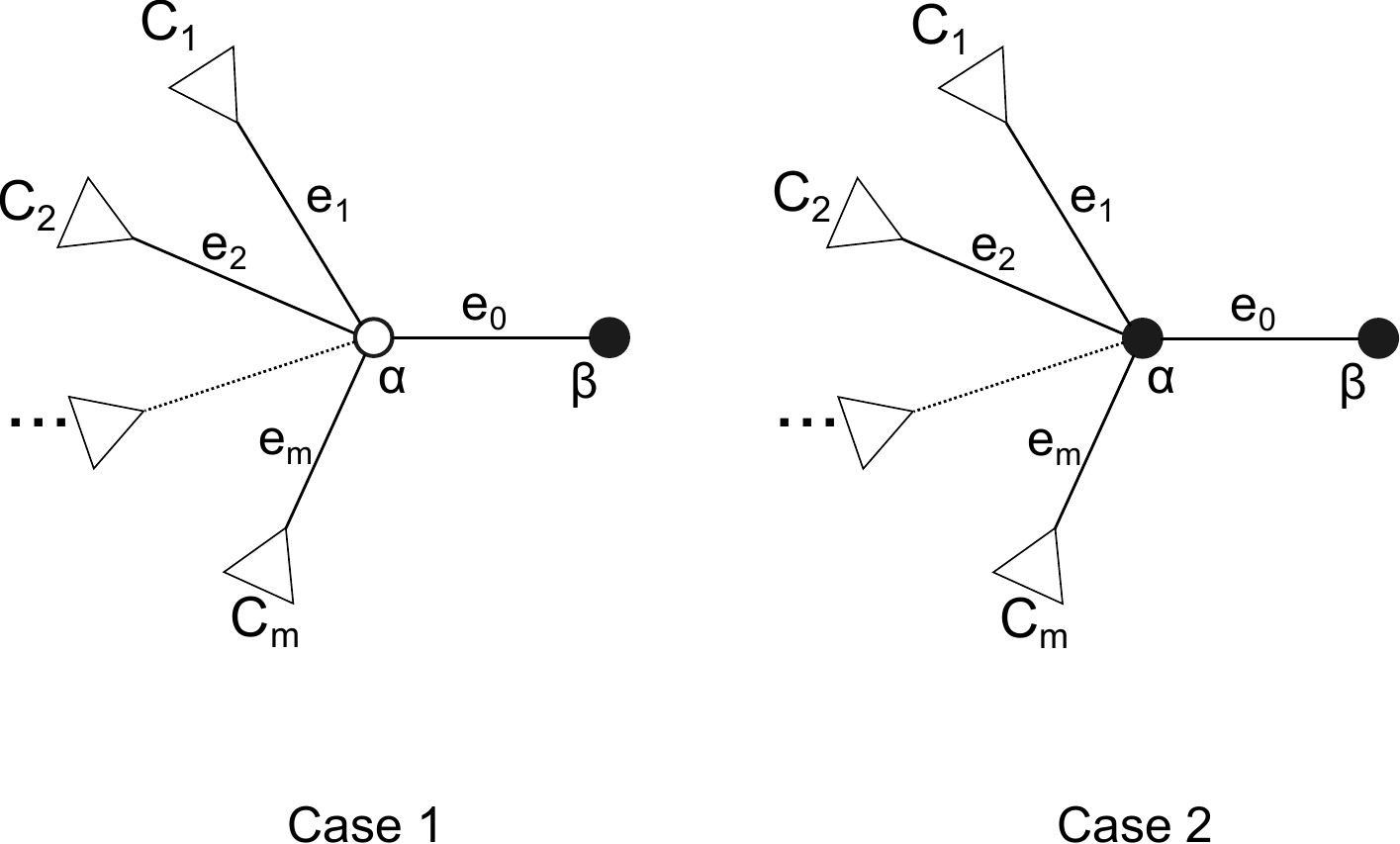}
\vspace *{1 em}
\caption{The two cases for the terminal edge $e_{0}$. $\alpha$ is labeled in case 1 and not labeled in case 2. The triangles represent subtrees.}
\label{fig:terminalEdges}
\end{figure}
There are two cases to consider for external branches, one if $\alpha$ is not labeled and the other if $\alpha$ is labeled see Fig. \ref{fig:terminalEdges}. In both cases the derivation of the branch length formula is similar to what has been described earlier and is omitted.

The branch length formulae turn out to be identical in all cases. The reader is referred to the supplementary material for the proof.
	
For a more detailed description of the algorithm for computing OLS branch lengths, the reader is referred to \citet{Bryant1997}.

Once $\mathbf{A}^{t}d$ has been computed, all branch lengths can be calculated in $O(n)$ time. Since there are $O(n)$ edges the time complexity of computing OLS branch lengths is $O(n^{2}$)

The overall time complexity of FJ is $O(n^{3})$. This can be reduced further if heuristics are used at the neighbor identification step, eqn. (\ref{eqn:neighborIdentificationStep}).

OLS branch lengths may be negative which has no biological interpretation. After estimating the branch lengths all branches that are shorter than $\epsilon$ and are incident to a latent vertex are contracted. If there is a branch between two labeled vertices that has a negative length, its length is set to $10^{-7}$. $10^{-7}$ is smaller than the smallest non-zero distance estimate computed in any of the simulation scenarios.

\subsection{Model selection}
Values of $\epsilon$ are inversely related to the number of latent vertices and thus inversely related to model complexity.

We performed model selection using three estimates for test error, cross-validation error, Akaike information criterion (AIC) and Bayesian information criterion (BIC). In all cases, model selection is performed by identifying the value of $\epsilon$ that minimizes the estimate for test error. Please refer to the Supplementary material for a description on how cross-validation error is computed.

AIC and BIC are Taylor series approximations of the Kullback-Leibler distance between the generative model which one wishes to recover and the model that is obtained by maximum likelihood estimation. These are formulated as,
$$\mbox{AIC} =  -2 \log \mbox{likelihood} + 2m$$
$$\mbox{BIC} =  -2 \log \mbox{likelihood} + m\log(n)$$

Under the likelihood framework, phylogenetic trees are probabilistic graphical models which are completely described by tree topology and branch lengths. $n$ denotes sample size and is given by sequence length. The number $m$ of parameters equals the number of branches in the tree.

We use FJ branch lengths as approximations of the maximum likelihood branch lengths. Likelihood is computed using Felsenstein's pruning algorithm which is a dynamic programming algorithm that enables efficient calculation of the likelihood \cite{Felsenstein1981}.

The calculation of cross-validation error is described on page 6 of the supplement.

\subsection{Related methods considered in the comparative validation}
\subsubsection{Sampled ancestors}
We used the sampled ancestors package \cite{Gavryushkina2014} of BEASTv2.3.0 \cite{Drummond2012} for the comparative validation of the FJ algorithm. The following models were considered: the GTR model for substitution, the four-category $\Gamma$ model for rate heterogeneity across sites, the strict molecular clock model and the fossilized birth death model for generating trees. Uniform priors were set for all model parameters. For all datasets, $10^8$ states were visited using Markov chain Monte Carlo (MCMC) and every $10^5th$ state was sampled. The first $5\%$ of the sampled states were discarded as burn-in and the effective sample size (ESS) was computed for all model parameters using the R package CODA \cite{Plummer2006}. ESS were found to be greater than 200 for all parameters across all the MCMC chains indicating that the chains were sufficiently long. The trees that are produced by BEASTv2.3.0 are rooted and contain the maximum number of latent vertices. The sampled trees were post-processed by unrooting them and contracting all terminal edges of length zero. We reported the average precision and recall of the post-processed sampled trees from the true tree.
 
\subsubsection{Recursive grouping and Chow-Liu recursive grouping}

For assessing the performance of RG and CLRG we used the Matlab implementation that was provided by the authors. Both of these methods are distance-based.
RG initially sets the active vertex set $V_{\text{a}}$ to the set of all labeled vertices. At each iteration $V_{\text{a}}$ is partitioned into so-called families using k-means clustering. For each family containing more than one vertex, a relationship test similar to the one used in FJ is performed. If there is a vertex that is the parent of all other vertices in the family then edges are added from the parent to each child. If no such parent is found then a latent vertex is introduced as the parent to all vertices of the family and corresponding edges are added. $V_{\text{a}}$ is reduced by removing all the children. This procedure is iterated until a connected graph is obtained.

CLRG starts by constructing a minimum spanning tree over all the labeled vertices. For each internal vertex $v_{i}$, the vertex set $V_{i}$ comprising of $v_{i}$ and its neighbors is constructed and RG is applied to distances between vertices in $V_{i}$, producing the tree $T_{i}$. The subgraph in the minimum spanning tree that is induced by $V_{i}$ is replaced by $T_{i}$. 

Both RG and CLRG require the setting of two thresholds, $\epsilon$ and $\tau$. The first threshold, $\epsilon$ is used for performing the relationship test. RG and CLRG additionally contract branches that are smaller than this threshold. The second threshold, $\tau$ is used to filter out large distances and only distances below this threshold are used when performing the relationship test. We optimized $\epsilon$ using BIC and set $\tau$ to a reasonably high value of $0.5$.

We modified the implementation provided by the authors, in order to correctly evaluate distances of value zero. Such distance estimates were encountered, predominantly, when the average branch length was the shortest and when a large fraction of internal vertices were labeled. The modification is that all distances of value zero were changed to $10^{-7}$.

\subsubsection{Neighbor-joining with edge contraction}
We implemented NJc in Python. NJc involves two steps. The first step is the construction of a tree using NJ. Subsequently all branches that are incident to a latent vertex and are smaller than a preselected threshold $\epsilon$ are contracted. We optimized $\epsilon$ using BIC.

\subsection{Acknowledgments}
PK is partially supported by German Center for Infection Research, grant no. DZIF 80008023. This work has been performed in the context of the EuResist Network GEIE, and the project MASTER-HIV/HEP which is funded by the German Health Ministry.
\subsection{Availability of code}
A program that constructs generally labeled trees using FJ-BIC is provided at https://bioinf.mpi-inf.mpg.de/publications/prabhavk/familyJoining.
\clearpage
\begin{center}
\Large{Supplementary material}
\end{center}
\section{Fast OLS for generally labeled trees}
In what follows we show that the branch length formula, eqn. (\ref{eqn:branchLengthEstimate}) (see also eqn. (10) in the main paper), that was derived by \citet{Bryant1997} for leaf-labeled trees is also applicable for generally labeled trees. We follow the same terminology that was defined in the main paper.

\begin{figure*}
  \centering
\includegraphics[width=\textwidth]{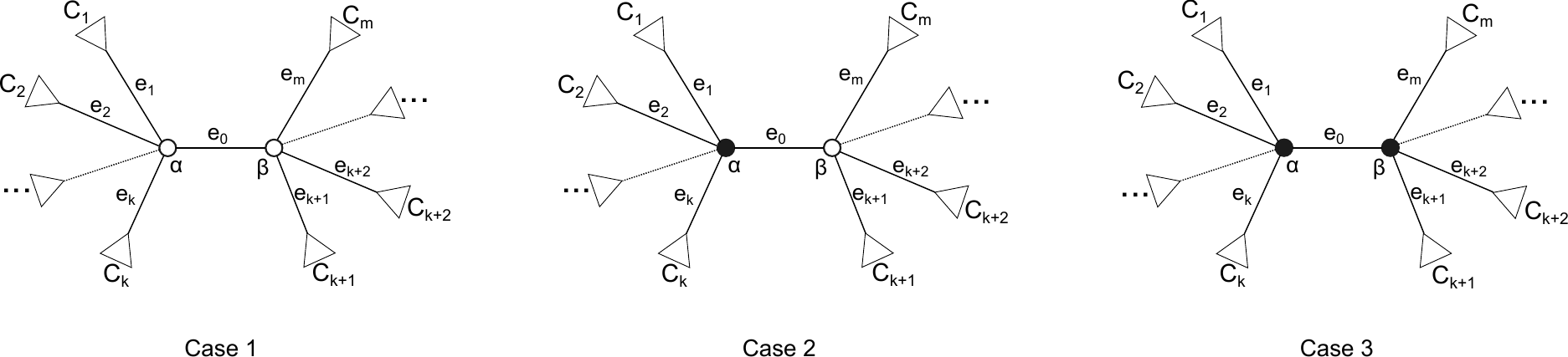}
\vspace*{0.2 em}
\caption{The three cases for the internal edge $e_{0}$. Case 1: Both $\alpha$ and $\beta$ are not labeled. Case 2: Only $\alpha$ is labeled. Case 3: Both $\alpha$ and $\beta$ are labeled. The triangles represent subtrees.}
\label{fig:internalEdges}
\end{figure*}

Consider the internal edge $e_{0}$ shown in Fig. \ref{fig:internalEdges} with adjacent edges $e_{1}, \ldots e_{k}, e_{k+1} \ldots e_{m}$. $e_{0}$ is incident to the vertices $\alpha$ and $\beta$. The respective sizes of the sides of the split defined by $e_{0}$ are $n_{\alpha}$ and $n_{\beta}$.

For each edge $e_{i}$, define $P_{i} = \sum_{x\in A_{i}, y \in B_{i}} p_{xy}$ where $A_{i}$ and $B_{i}$ are the sides of the split defined by edge $e_{i}$. Here $p_{xy}$ denotes the length of the path from $x$ to $y$ when branch lengths are determined by OLS. It turns out that $P_{i} = \delta_{i}^{T}d$.

For each edge $e_{i}$, $i \neq 0$, let $C_{i}$ be the side of the split defined by $e_{i}$ that does not contain $\alpha$ and $\beta$. $n_{i}$ is the cardinality of $C_{i}$. Define
\begin{equation*}
  Q_{i}=\begin{cases}
    \sum_{x \in C_{i}} p_{\alpha x}, & \mbox{if } 1 \leq i \leq k\\
    \sum_{x \in C_{i}} p_{\beta x}, & \mbox{if } k+1 \leq i \leq m\\
  \end{cases}
\end{equation*}
If both $\alpha$ and $\beta$ are not labeled (Case 1 in Fig. \ref{fig:internalEdges}) it can be shown that \citep{Bryant1997}
$$\underline{P} = (nI - 2N)\underline{Q} +NU\underline{Q} + b_{0}N\underline{v}$$
where $N$ is the $m \times m$ diagonal matrix with $(n_{1},n_{2},\ldots, n_{m})$ on the diagonal, $I$ is the identity matrix, $\underline{Q} = (Q_{1},Q_{2},\ldots,Q_{m})^{T}$, $U$ is the $m \times m$ matrix of ones, $\underline{v}$ is the vector with $n_{\beta}$ in positions $1$ to $k$ followed by $n_{\alpha}$ in positions $k+1$ to $m$, $\underline{P} = (P_{1},P_{2},\ldots,P_{m})^{T}$, $n$ is the total number of labeled vertices, and $b_{0}$ is the branch length of the edge $e_{0}$ 

Similarly for the internal edge $e_{0}$,
\begin{equation*}
P_{0}=\underline{v}^{T}\underline{Q}+ n_{\alpha}n_{\beta}b_{0}
\end{equation*}
Letting $X = (nN^{-1}-2I+U)$ and substituting $\underline{Q}$ gives the following branch length estimate.
\begin{equation*}
b_{0} = \dfrac{P_{0}-\underline{v}^{T}X^{-1}N^{-1}\underline{P}}{n_{\alpha}n_{\beta}-\underline{v}^{T}X^{-1}\underline{v}}
\end{equation*}
For cases where only $\alpha$ and both $\alpha$ and $\beta$ are labeled, respectively, the derivation of the equations are similar to that described in \citet{Bryant1997} and is described below.

\subsection*{Case 2: $\alpha$ is labeled and $\beta$ is not labeled}
For edges $e_{i}$ incident to $\alpha$, $i = 1\ldots k$, we have
\begin{align*}
P_{i}  &= \sum_{x\in A_{i}}\sum_{y \in B_{i}} p_{xy} \\
&=\sum_{j = 1, j\neq i}^{m}\sum_{x\in C_{i}}\sum_{y\in C_{j}} p_{xy} + \sum_{x\in C_{i}}p_{\alpha x}\\
&=\sum_{j = 1, j\neq i}^{k}\sum_{x\in C_{i}}\sum_{y\in C_{j}}(p_{\alpha x} + p_{\alpha y}) + \sum_{j = k+1}^{m}\sum_{x\in C_{i}}\sum_{y\in C_{j}}(p_{\alpha x} + b_{0} + p_{\beta y}) + \sum_{x\in C_{i}}p_{\alpha x}\\
&= \sum_{j = 1,j \neq i }^{k}\!\!\!\![n_{j}Q_{i} + n_{i}Q_{j}] + \sum_{j = k+1}^{m}\!\!\![n_{j}Q_{i} + n_{i}Q_{j} + n_{i}n_{j}b_{0}] + Q_{i}\\
&= (n-n_{i}-1)Q_{i} + n_{i}(Q_{1}+\ldots+Q_{i-1}+Q_{i+1}+\ldots+Q_{m}) + n_{i}n_{\beta}b_{0} + Q_{i}\\
&=(n-2n_{i})Q_{i} + n_{i}\sum_{j=1}^{m}Q_{j} + n_{i}n_{\beta}b_{0}
\end{align*}
For edges $e_{i}$ incident to $\beta$, $i = k+1\ldots m$, we have
\begin{align*}
P_{i}  &= \sum_{x\in A_{i}}\sum_{y \in B_{i}} p_{xy} \\
&=\sum_{j = 1, j\neq i}^{m}\sum_{x\in C_{i}}\sum_{y\in C_{j}} p_{xy} + \sum_{x\in C_{i}}p_{\alpha x}\\
&=\sum_{j = 1}^{k}\sum_{x\in C_{i}}\sum_{y\in C_{j}}(p_{\beta x} + b_{0} +p_{\alpha y}) + \sum_{j = k+1, j\neq i}^{m}\sum_{x\in C_{i}}\sum_{y\in C_{j}}(p_{\beta x} + p_{\beta y}) + \sum_{x\in C_{i}}(p_{\beta x}+b_{0})\\
&= (\sum_{j = 1}^{k}n_{j}Q_{i} + n_{i}Q_{j} +n_{i}n_{j}b_{0}) + (\sum_{j = k+1,j \neq i }^{m}n_{j}Q_{i} + n_{i}Q_{j}) + Q_{i} + n_{i}b_{0}\\
&= (n-n_{i}-1)Q_{i} + n_{i}(Q_{1}+\ldots+Q_{i-1}+Q_{i+1}+\ldots+Q_{m}) + n_{i}(n_{\alpha}-1)b_{0} + Q_{i}+n_{i}b_{0}\\
&=(n-2n_{i})Q_{i} + n_{i}\sum_{j=1}^{m}Q_{j} + n_{i}n_{\alpha}b_{0}
\end{align*}

In matrix form,
\begin{align*}
&\underline{P} = (nI - 2N)\underline{Q} +NU\underline{Q} + b_{0}N\underline{v}\\
&\Leftrightarrow N(nN^{-1}-2I+U)\underline{Q} = \underline{P} - b_{0}N\underline{v}
\end{align*}

 Setting $X = (nN^{-1}-2I+U)$ and rearranging, we get
$$\underline{Q} = X^{-1}N^{-1}\underline{P}-b_{0}X^{-1}\underline{v}$$

For the internal edge $e_{0}$ we have
\begin{align*}
P_{0} &= \sum_{i=1}^{k}\sum_{j=k+1}^{m}\sum_{x \in C_{i}, y \in C_{j}} p_{xy} + \sum_{j=k+1}^{m} \sum_{x \in C_{j}}(b_{0} + p_{\beta x})\\
&=(\sum_{i=1}^{k}\sum_{j=k+1}^{m}\sum_{x \in C_{i}, y \in C_{j}}p_{\alpha x} + b_{0} + p_{\beta y}) + n_{\beta}b_{0} + \sum_{j=k+1}^{m}Q_{j}\\
&=(\sum_{i=1}^{k}\sum_{j=k+1}^{m}n_{j}Q_{i} + n_{i}n_{j}b_{0} + n_{i}Q_{j}) + n_{\beta}b_{0} + \sum_{j=k+1}^{m}Q_{j}\\
&=\sum_{i=1}^{k}n_{\beta}Q_{i} + \sum_{j=k+1}^{m}(n_{\alpha}-1)Q_{j} + (n_{\alpha}-1)n_{\beta}b_{0} +  n_{\beta}b_{0} + \sum_{j=k+1}^{m}Q_{j}\\
&=\underline{v}^{T}\underline{Q}+ n_{\alpha}n_{\beta}b_{0}
\end{align*}

After substituting \underline{Q} and rearranging we get,
\begin{equation}
\label{eqn:branchLengthEstimate}
b_{0} = \dfrac{P_{0}-\underline{v}^{T}X^{-1}N^{-1}\underline{P}}{n_{\alpha}n_{\beta}-\underline{v}^{T}X^{-1}\underline{v}}
\end{equation}

\subsection*{Case 3: Both $\alpha$ and $\beta$ are labeled}
For edges $e_{i}$ incident to $\alpha$, $i = 1\ldots k$, we have
\begin{align*}
P_{i}  &= \sum_{x\in A_{i}}\sum_{y \in B_{i}} p_{xy} \\
&=\left[\sum_{j = 1, j\neq i}^{m}\sum_{x\in C_{i}}\sum_{y\in C_{j}} p_{xy}\right] + \sum_{x\in C_{i}}p_{\alpha x} + \sum_{x\in C_{i}}p_{\beta x}\\
&=\left[\sum_{j = 1, j\neq i}^{k}\sum_{x\in C_{i}}\sum_{y\in C_{j}}p_{\alpha x} + p_{\alpha y}\right] + \left[\sum_{j = k+1}^{m}\sum_{x\in C_{i}}\sum_{y\in C_{j}}p_{\alpha x} + b_{0} + p_{\beta y}\right] + 2\sum_{x\in C_{i}}p_{\alpha x} + n_{i}b_{0}\\
&= \left[\sum_{j = 1,j \neq i }^{k} n_{j}Q_{i} + n_{i}Q_{j}\right] + \left[\sum_{j = k+1}^{m} n_{j}Q_{i} + n_{i}Q_{j} + n_{i}n_{j}b_{0}\right] + 2Q_{i}+n_{i}b_{0}\\
&= (n-n_{i}-2)Q_{i} + n_{i}(Q_{1}+\ldots+Q_{i-1}+Q_{i+1}+\ldots+Q_{m}) + n_{i}b_{0}(1+\!\!\!\!\sum_{j=k+1}^{m}\!\!\!\!n_{j}) + 2Q_{i}\\
&=(n-2n_{i})Q_{i} + n_{i}\sum_{j=1}^{m}Q_{j} + n_{i}n_{\beta}b_{0}
\end{align*}

By symmetry, for edges $e_{i}$ incident to $\beta$, $i = k+1\ldots m$, we have,
$$P_{i}=(n-2n_{i})Q_{i} + n_{i}\sum_{j=1}^{m}Q_{j} + n_{i}n_{\alpha}b_{0}$$

In matrix form,
$$\underline{P} = (nI - 2N)\underline{Q} +NU\underline{Q} + b_{0}N\underline{v}$$

For the internal edge $e_{0}$ we have

\begin{align*}
P_{0} &= \sum_{i=1}^{k}\sum_{j=k+1}^{m}\sum_{x \in C_{i}, y \in C_{j}} p_{xy} + \left[\sum_{j=1}^{k} \sum_{x \in C_{j}}b_{0} + p_{\alpha x}\right] + \left[\sum_{j=k+1}^{m} \sum_{x \in C_{j}}b_{0} + p_{\beta x}\right] + b_{0}\\
&=\left[\sum_{i=1}^{k}\sum_{j=k+1}^{m}\sum_{x \in C_{i}, y \in C_{j}}\!\!\!\!p_{\alpha x} + b_{0} + p_{\beta y}\right] + (n_{\alpha}+n_{\beta}-1)b_{0} + \sum_{j=1}^{m}Q_{j}\\
&=\left[\sum_{i=1}^{k}\sum_{j=k+1}^{m}\!\!n_{j}Q_{i} + n_{i}n_{j}b_{0} + n_{i}Q_{j}\right] + (n_{\alpha}+n_{\beta}-1)b_{0} + \sum_{j=1}^{m}Q_{j}\\
&=(n_{\beta}-1)\sum_{i=1}^{k}Q_{i} + (n_{\alpha}-1)\sum_{j=k+1}^{m}Q_{j} + (n_{\alpha}-1)(n_{\beta}-1)b_{0} +(n_{\alpha}+n_{\beta}-1)b_{0} + \sum_{j=1}^{m}Q_{j}\\
&=n_{\beta}\sum_{i=1}^{k}Q_{i} + n_{\alpha}\sum_{i=k+1}^{m}Q_{i} + n_{\alpha}n_{\beta}b_{0}\\
&=\underline{v}^{T}\underline{Q}+ n_{\alpha}n_{\beta}b_{0}
\end{align*}
After substituting \underline{Q} and rearranging we get,
$$b_{0} = \dfrac{P_{0}-\underline{v}^{T}X^{-1}N^{-1}\underline{P}}{n_{\alpha}n_{\beta}-\underline{v}^{T}X^{-1}\underline{v}}$$

\begin{figure}
\centering
\includegraphics[width=0.45\textwidth]{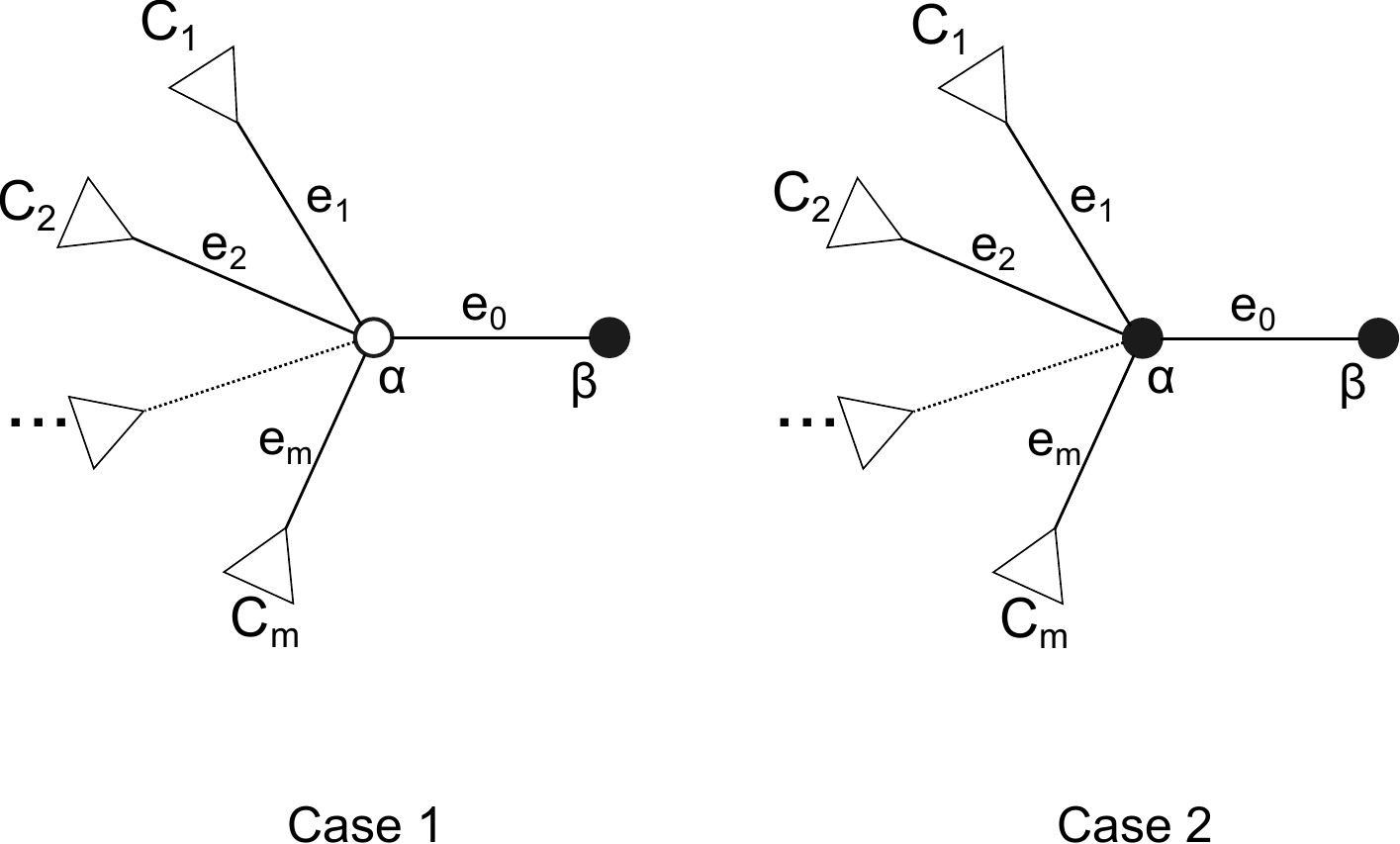}
\vspace *{1 em}
\caption{The two cases for the terminal edge $e_{0}$. $\alpha$ is labeled in case 1 and not labeled in case 2. The triangles represent subtrees.}
\label{fig:terminalEdges}
\end{figure}

Consider the terminal edge $e_{0}$ shown in Fig. \ref{fig:terminalEdges} with adjacent edges $e_{1},e_{2}\ldots e_{m}$. $e_{0}$ is incident to the vertices $\alpha$ and $\beta$. The respective sizes of the sides of the split defined by $e_{0}$ are $n_{\alpha}$ and $n_{\beta}$. Since $e_{0}$ is a terminal edge the leaf $\beta$ is labeled. There are two cases to consider depending on if $\alpha$ is labeled or not labeled.

If $\alpha$ is not labeled (Case 1 in Fig. \ref{fig:terminalEdges}), the branch length formula given by \citet{Bryant1997} is
$$b_{0} = \dfrac{P_{0}-\underline{v}^{T}X^{-1}N^{-1}\underline{P}}{n_{\alpha}n_{\beta}-\underline{v}^{T}X^{-1}\underline{v}}$$
where  $n_\alpha = (n-1)$, $n_\beta = 1$ and $k=m$.
If $\alpha$ is labeled (Case 2 in Fig. \ref{fig:terminalEdges}), the branch length formula can be derived as follows.

For edges $e_{i}$ incident to $\alpha$ we have,
\begin{align*}
P_{i} &= \sum_{x\in A_{i}}\sum_{y \in B_{i}} p_{xy}\\
&=\sum_{j = 1, j\neq i}^{m}\sum_{x\in C_{i}}\sum_{y\in C_{j}} p_{xy} + \sum_{x\in C_{i}}(p_{\alpha x} + p_{\beta x})\\
&=\sum_{j = 1, j\neq i}^{m}\sum_{x\in C_{i}}\sum_{y\in C_{j}} (p_{\alpha x} + p_{\alpha y}) + \sum_{x\in C_{i}}(2p_{\alpha x} + b_{0})\\
&=\sum_{j = 1, j\neq i}^{m}[n_{j}Q_{i}+n_{i}Q_{j}] +2Q_{i}+n_{i}b_{0}\\
&=(n-n_{i}-2)Q_{i} + n_{i}\sum_{j = 1, j\neq i}^{m}Q_{j}+2Q_{i}+n_{i}b_{0}\\
&=(n-2n_{i})Q_{i} + n_{i}\sum_{j = 1}^{m}Q_{j}+n_{i}b_{0}\\
\end{align*}
In matrix form,
$$\underline{P} = (nI - 2N)\underline{Q} +NU\underline{Q} + b_{0}N\underline{v}$$
For the terminal edge $e_0$ we have,
\begin{align*}
P_{0} &= \sum_{i=1}^{m}\sum_{x \in C_{i}} p_{\beta x} + b_{0}\\
&=(\sum_{i=1}^{m}\sum_{x \in C_{i}} p_{\alpha x} +b_{0}) + b_{0}\\
&=\sum_{i=1}^{m}Q_{i} + (n-1)b_{0}\\
&=\underline{v}^{T}\underline{Q}+ n_{\alpha}n_{\beta}b_{0}
\end{align*}
where  $n_\alpha = (n-1)$, $n_\beta = 1$ and $k=m$.

After substituting \underline{Q} and rearranging we get,
$$b_{0} = \dfrac{P_{0}-\underline{v}^{T}X^{-1}N^{-1}\underline{P}}{n_{\alpha}n_{\beta}-\underline{v}^{T}X^{-1}\underline{v}}$$
\pagebreak
\section{Molecular clock rate inferred by SA}
\begin{figure}[htbp]
\centering
\includegraphics[width=0.8\textwidth]{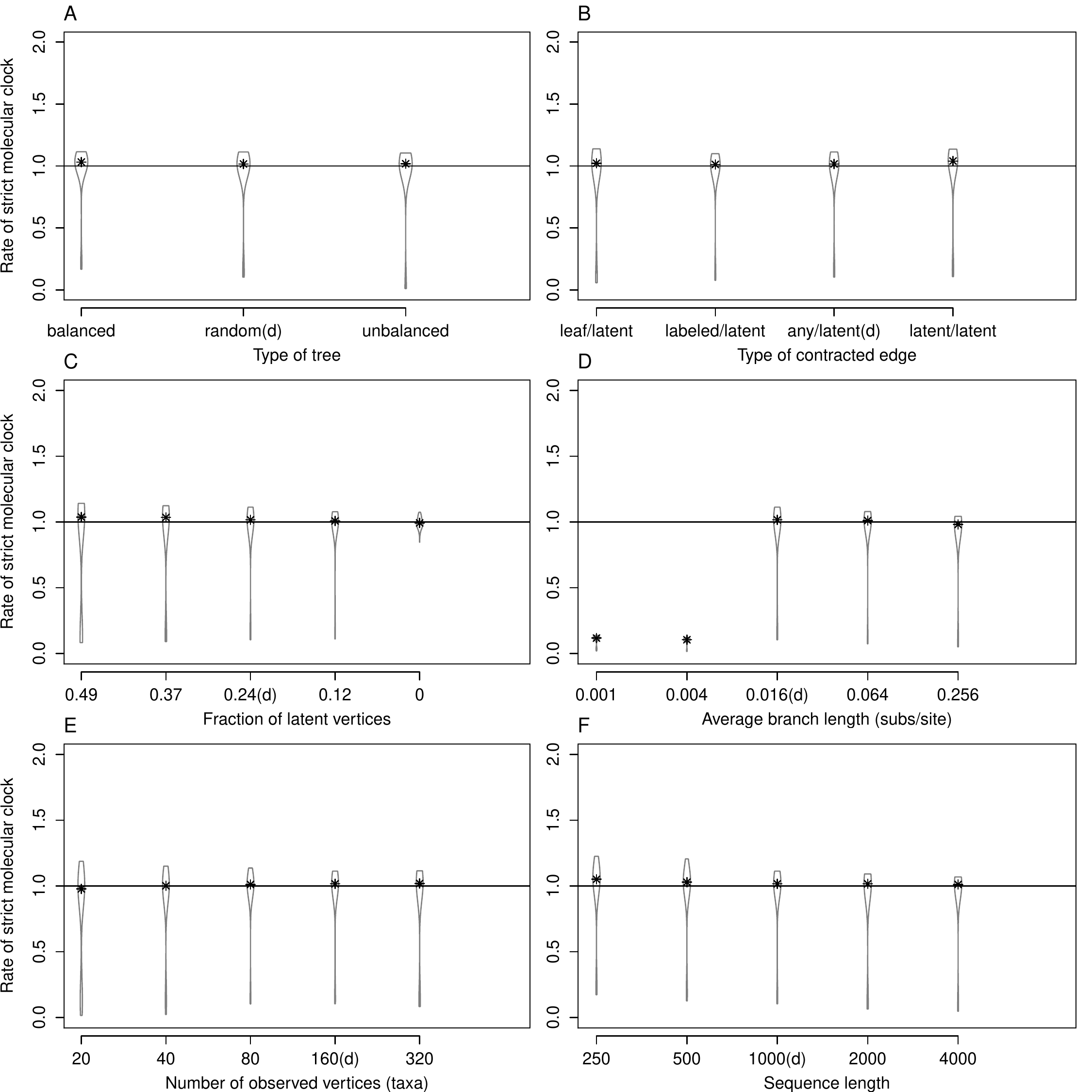}
\caption{Rate of the strict molecular clock that is estimated by SA. The true rate of the strict molecular clock is 1.0 subs./site/time in all simulation scenarios.}
\end{figure}

\section{Comparison of various FJ-based methods}
For computing cross-validation error the original sequence alignment with $L$ columns was partitioned into $K$ validation alignments by randomly sampling $L/K$ columns without replacement. For each validation alignment, the corresponding training alignment was constructed using the complimentary set of $L-L/K$ alignment columns. This procedure was repeated $R$ times, giving $RK$ training and validation alignments in total. ML distances were computed for all training and validation alignments. For a fixed value of $\epsilon$, FJ trees were constructed for each training distance matrix. We set $R$ to 10 and tried two values for $K$, i.e., 3 and 5. Test error was computed as the residual sum of squares between the fitted distances (path length on the tree) and the corresponding distances computed from the validation alignment. We then found the $\epsilon$ that minimized expected test error as this would yield the most generalizable model.
\[\arg\min\limits_{\epsilon} \displaystyle\sum_{k}\displaystyle\sum_{i,j}\!\!\!\underbrace{({d_{T(\!\epsilon,k\!)}}(i,j)}_{\text{distance in fitted tree}}-\underbrace{{d_{V(k)}}(i,j))^{2}}_{\text{distance in validation set}} \]
where $T(\epsilon,k)$ is the tree constructed at threshold $\epsilon$ using distances from the $k^{\text{th}}$ training alignment and $V(k)$ is the $k^{\text{th}}$ validation alignment.
\begin{figure}[htbp]
\centering
\includegraphics[width=0.8\textwidth]{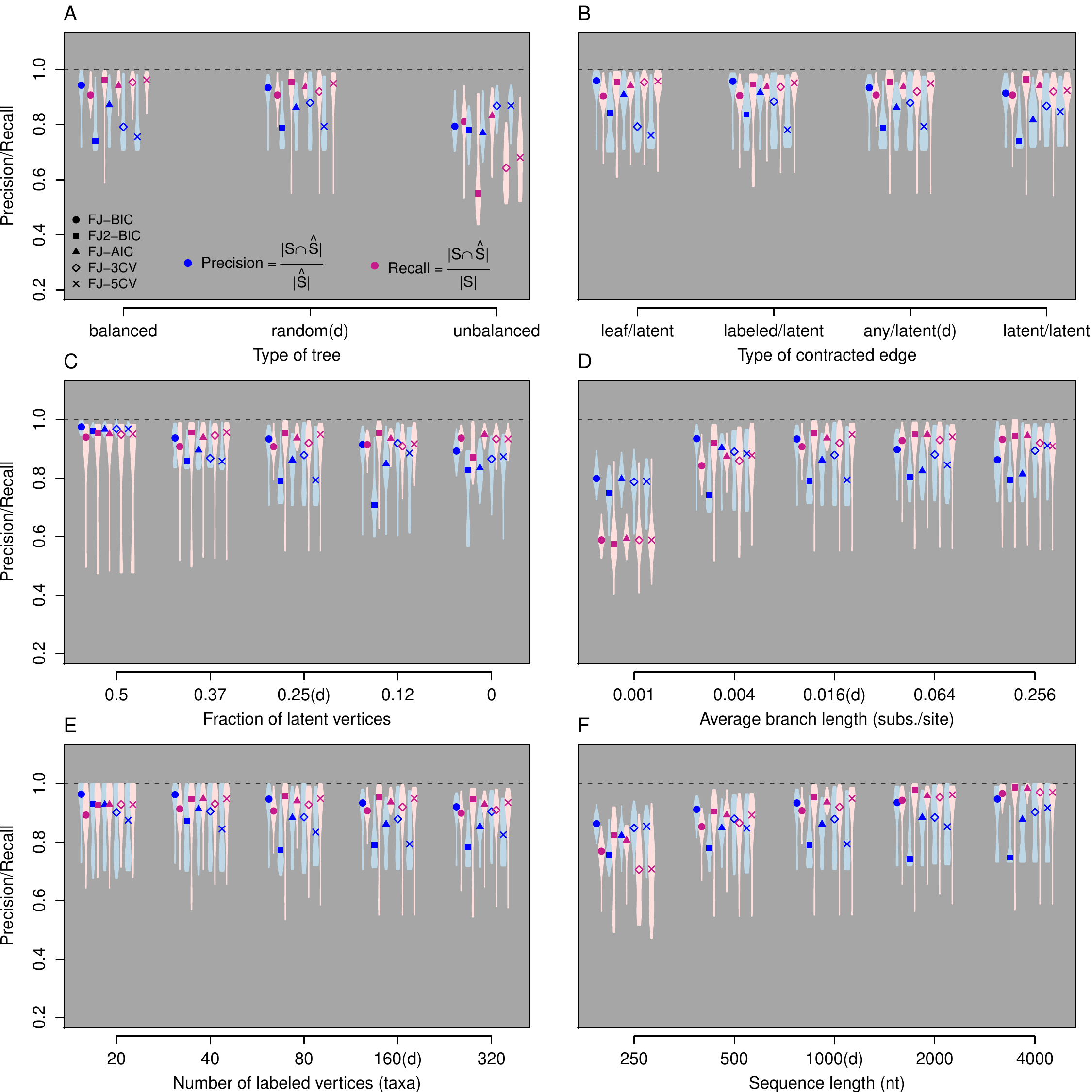}
\caption{A comparison of various FJ-based methods. FJ-BIC is the method that is presented in the main paper. FJ2-BIC checks if siblings have a parent using the criterion shown in eqn. (4) of the main paper. FJ-AIC uses AIC for model selection. FJ-3CV and FJ-5CV performs model selection using 3-fold CV and 5-fold CV respectively.}
\label{fig:precisionRecall_FJ}
\end{figure}

\bibliographystyle{natbib}
\bibliography{FamilyJoining}

\end{document}